\documentclass[preprint,12pt]{elsarticle}

\usepackage{amssymb}
\usepackage{amsfonts}
\usepackage{amsmath}
\usepackage{amsthm}
\usepackage[toc]{appendix}
\usepackage{color}
\usepackage{graphicx} 
\usepackage{subfigure}
\usepackage{multirow}
\usepackage{float}
\usepackage{nicefrac}
\usepackage{verbatim}
\usepackage{tabularx}
\usepackage{booktabs}
\usepackage{times}
\usepackage{balance}
\usepackage{eurosym}
\usepackage{enumitem}
 \usepackage{amsthm}

\journal{Computer Communications}

\begin{document}

\begin{frontmatter}



\title{Unsatisfied Today, Satisfied Tomorrow: a simulation framework for performance evaluation of crowdsourcing-based network monitoring}

\author{Andrea Pimpinella, Marianna Repossi, Alessandro E. C. Redondi\\ 
Dipartimento di Elettronica, Informazione e Bioingegneria, \\
Politecnico di Milano, Italy\\
e-mail: \{andrea.pimpinella, marianna.repossi, alessandroenrico.redondi\}@polimi.it \\
}


\begin{abstract}
Network operators need to continuosly upgrade their infrastructures in order to keep their customer satisfaction levels high. Crowdsourcing-based approaches are generally adopted, where customers are directly asked to answer surveys about their user experience. Since the number of collaborative users is generally low, network operators rely on Machine Learning models to predict the satisfaction levels/QoE of the users rather than directly measuring it through surveys. Finally, combining the true/predicted user satisfaction levels with  information on each user mobility (e.g, which network sites each user has visited and for how long), an operator may reveal critical areas in the networks and drive/prioritize investments properly.
In this work, we propose an empirical framework tailored to assess the quality of the detection of under-performing cells starting from subjective user experience grades. The framework allows to simulate diverse networking scenarios, where a network characterized by a small set of under-performing cells is visited by heterogeneous users moving through it according to realistic mobility models. The framework simulates both the processes of satisfaction surveys delivery and users satisfaction prediction, considering different delivery strategies and evaluating prediction algorithms characterized by different prediction performance. We use the simulation framework to test empirically the performance of under-performing sites detection in general scenarios characterized by different users density and mobility models to obtain insights which are generalizable and that provide interesting guidelines for network operators.
\end{abstract}


\end{frontmatter}



\section{Introduction}
\label{sec:introduction}
According to recent Cisco estimates \cite{Ciscoreport}, by 2021 mobile cellular networks will connect more than 11 billion mobile devices and will be responsible for more than one fifth of the total IP traffic generated worldwide. Moreover, the global average broadband speed will more than double from 2018 to 2023, from 45.9 Mbps to 110.4 Mbps. This will result in increased utilisation of high-bandwidth demanding applications, such as on-demand 4K video streaming, cloud storage, etc.
To face this unprecedented growth in both volume of mobile traffic and data rate needs of customers, network operators continuously invest in all network domains, including but not limited to spectrum, radio access network (RAN) infrastructure, transmission and core networks. The final goal of such investments is to generate profit by (i) attracting as many customers as possible and (ii) minimizing the number of churners, i.e., users who stop their current subscriptions and move to a different operator.

Concerning the latter point, a well established process mobile operators perform to avoid churns is to monitor their customers satisfaction levels through directed surveys: as an example, the Net Promoter Score (NPS) survey asks users to indicate the likelihood of recommending the network operator to a friend or colleague on a scale from 0 to 10. In addition to such a generic survey, operators often ask customers to reply very specific questions related to the user satisfaction or Quality of Experience (QoE) relative to certain network services (network coverage, voice and video quality, etc.), which can better highlight possible problems in the network, such as under-perfoming or malfuctioning network cells/sites. Unfortunately, such a direct way to track users satisfaction is costly and cumbersome for operators, mainly due to the generic poor cooperative attitude of customers. Moreover, the problem of the reliability of users' replies to such surveys is subject to intense investigations \cite{raykar2010learning,karger2011iterative,hossfeld2011memory, shah2015approval}: regardless of the subject of the surveys, studies confirm that it is not a trivial task to gather reliable responses from crowds, especially when no reward systems are conceived.

To cope with these issues, several studies in the recent literature addressed the problem of predicting the satisfaction level of customers, rather than directly measuring it through surveys \cite{huang2015telco, tong2017research, swetha2018evaluation, boz2019mobile, pimpinella2019towards}. Following the renovated interest in big data, machine learning and artificial intelligence, the goal of such works is to identify the set of unsatisfied customers starting from a large variety of objective features, both operative (e.g., average throughput and signal quality) and business-related (e.g., gender, age or tariff plan). Such features, and the corresponding ground-truth satisfaction level, are generally used to train machine-learning models, eventually being able to estimate the satisfaction levels/QoE of a much larger population. Finally, combining the true/predicted users satisfaction levels with  information on each user mobility (e.g, which network sites each user has visited and for how much time), an operator may reveal critical areas in the networks and drive/prioritize investments properly.

However, the detection of under-performing cells starting from true/predicted subjective grades has its own issues. First, users are heterogeneous and their perception of network quality is highly subjective. Second, when a negative satisfaction expressed by a user refers to a long period of time (e.g., one month), it is difficult to identify which of the network sites visited during that period is the most responsible. Third, in case the user satisfaction level is estimated through a machine learning algorithm, a prediction error is likely to be expected. Therefore, in this complex scenario, an operator may argue about the validity/quality of the detected under-performing cells. 
To solve these issues, we propose an empirical framework tailored to assess the quality of the detection of under-performing cells starting from subjective user grades. In details, the contributions of this paper are:
\begin{enumerate}
	\item We build a framework that allows to simulate a network composed of a (small) set of under-performing/malfunctioning cells, with heterogeneous users moving freely in it according to realistic mobility models. Depending on each user mobility and subjective profile, the framework allows to obtain each user's (true) satisfaction level.
	\item The framework also simulates the process of satisfaction surveys delivery performed by the operator, which is able to sample only a subset of the true user satisfaction levels through surveys. We consider two different delivery strategies: a completely random one and one which maximizes the number of covered network sites.
	\item Moreover, the proposed framework allows to simulate the process of users satisfaction prediction using a machine learning algorithm whose performance can be changed at will. This allows to quantify the impact of prediction errors on the detection process, and to understand what are the minimum performance a prediction model for user satisfaction should possess to be applied in the overall methodology. 
	\item Finally, we test empirically the simulation framework with different users density and mobility models to obtain insights which are generalizable.
\end{enumerate} 

The remainder of this article is organized as follows: Section \ref{sec:system} describes a general crowdsourcing network monitoring process than can be adopted by an operator to perform detection of under-performing sites, leveraging both objective data and true/predicted user satisfaction levels; Section \ref{sec:framework} describes the simulation framework that can be used to assess the quality of the detection process. Diverse scenarios characterized by different users density, mobility models and surveys delivery strategies are simulated in Section 4, to empirically test the performance of the detection of under-performing cellular sites. Section \ref{sec:sota} reviews the relevant literature on QoE prediction and QoE-based issues detection in cellular networks. Finally, Section \ref{sec:conclusion} summarises remarks and conclusions.
\section{Under-Performing Sites Detection Process}\label{sec:system}
Detecting possible issues in an operator network infrastructure using information about the perceived user experience is a process known under the name of crowdsourcing network monitoring, a field which has received increasing attention in the last few years ~\cite{choffnes2010crowdsourcing, faggiani2014smartphone, ren2015exploiting}. According to this approach, the mobile operator administers to its customers population $\mathcal{U}, |\mathcal{U}|=N$  a set of user experience/satisfaction surveys (either directly or through the help of proper apps installed on the user equipments), whose answers may help to reveal critical/under-performing network sites, hence steering investments in the right directions (e.g., increasing the bandwidth or the output power available at specific base stations). Rather than detecting all sites responsible for users dissatisfaction, a more convenient output for a mobile operator consists in a \textit{site ranking}, i.e., a sorted list of network sites in which the ones responsible for the highest number of unsatisfied users appear at the top positions. In such a way, an operator may allocate the available budget for investment to the first $k$ sites in the list in a prioritized fashion. 

\begin{figure}[t]
    \centering
    \includegraphics[width=\columnwidth]{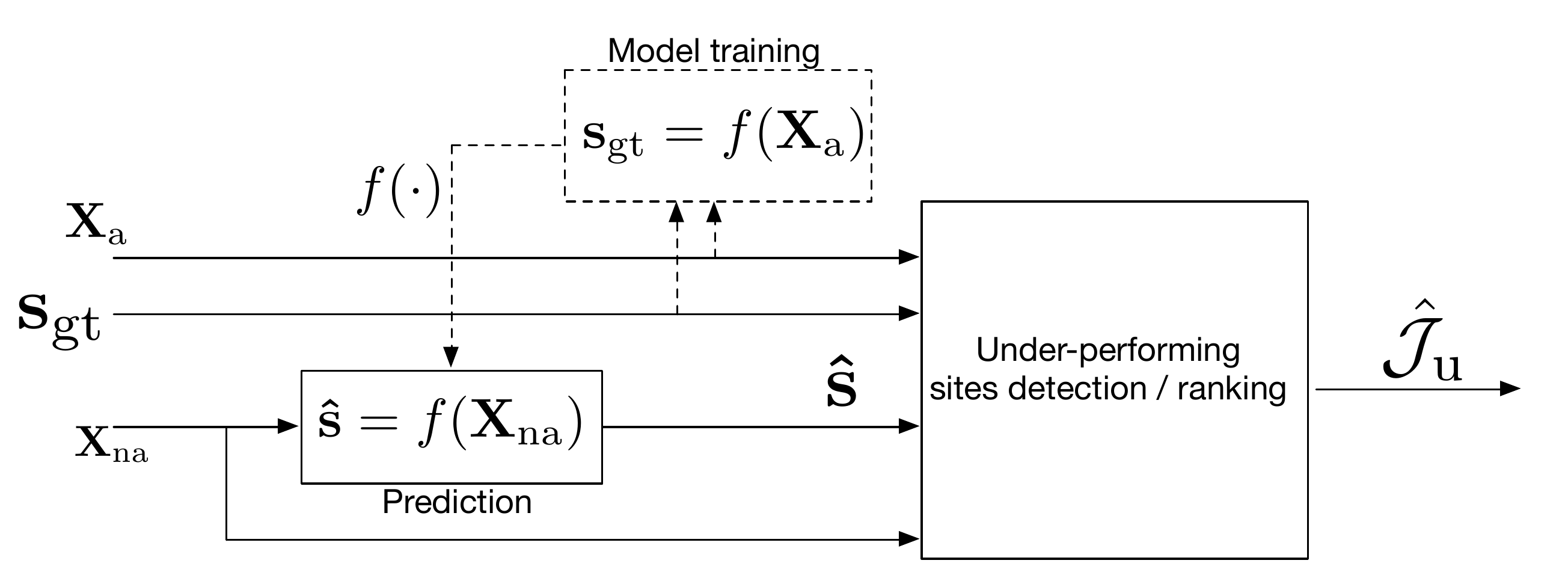}
    \caption{General process for crowdsourcing-based site ranking. The satisfaction grades from the users, true or predicted, are combined with objective information such as user visit times to detect critical network sites and rank them according to their impact on user satisfaction. Dashed lines refer to the fact that the model $f(\cdot)$ is independent from the detection process and may be updated asynchronosuly whenever new survey responses are gathered by the operator.}
    \label{fig:anomalydetector}
\end{figure}

When the responses gathered from the users are few (and this is often the case \cite{pimpinella2019towards}), operators may rely on data science techniques to predict the satisfaction of additional users, artificially enlarging the set of available responses. This is generally obtained by exploiting pre-trained machine learning  models that correlate objective network measurements collected from the users (e.g., throughput, channel quality, amount of time spent with limited service) with the users perceived satisfaction. Since the objective network measurements are generally available for a much larger amount of users compared to the (subjective) satisfaction responses, this strategy allows to greatly enlarge the knowledge base usable for detecting or ranking under-performing sites. The general process is illustrated in Figure \ref{fig:anomalydetector}: let $\mathcal{U} = \{\mathcal{U}_\text{a} \cup \mathcal{U}_\text{na}\}$ be the total set of network users, composed of customers whose survey response is available ($\mathcal{U}_\text{a}$) or not ($\mathcal{U}_\text{na}$). Similarly, let $\mathbf{X}_\text{a}$ and $\mathbf{X}_\text{na}$ be the set of objective network measurements for the two sets of users. A machine learning model $f(\cdot)$, trained and possibly updated with the knowledge coming from users whose answers $\mathbf{s}\textsubscript{gt}$ are available, can be used to predict the satisfactions $\mathbf{\hat{s}}$ of non-answering users. We underline that the model $f(\cdot)$ can be trained independently of the detection process and updated any time new surveys responses are gathered by the operator. Finally, the objective network measurements ($\mathbf{X}_\text{a}$, $\mathbf{X}_\text{na}$) and the true and predicted user satisfaction ($\mathbf{s}\textsubscript{gt}$ and $\mathbf{\hat{s}}$) are leveraged to produce a ranked list of sites in the network, which we refer to as $\mathcal{\hat{J}_\text{u}}$.

Several important questions related to such an approach can be raised by a mobile network operator: 
\begin{enumerate}[label=Q.\arabic*]
	\item \label{Q1} \textit{Ranking strategy}: Assuming the availability of the (true) satisfaction of the entire set of users, how can under-performing sites be ranked/detected? 
	\item \label{Q2} \textit{User heterogeneity:} Different users react to network issues in different way. How does such heterogeneity impact on the detection of under-performing sites?  
	\item \label{Q3} \textit{Prediction errors:} When a ML model $f(\cdot)$ is used to predict the satisfaction of the non-answering users, a prediction error is generally expected. How does such an error impact on the ranking/detection of under-performing sites? 
	\item \label{Q4} \textit{Users density:} What is the relationship among the cardinality of the sets of answering and non-answering users, the number of sites in the network and the performance of the ranking/detection operation?
	\item \label{Q5} \textit{Survey delivery:} If only a subset of users is expected to answer the satisfaction surveys, is there a way to select such a subset in order to increase the performance of the detection process? 
\end{enumerate}
In the following, we describe a simulation framework that an operator can leverage in order to find answers to such questions. 


\section{Simulation Framework} \label{sec:framework}
\begin{figure}[t]
    \centering
    \includegraphics[width=\columnwidth]{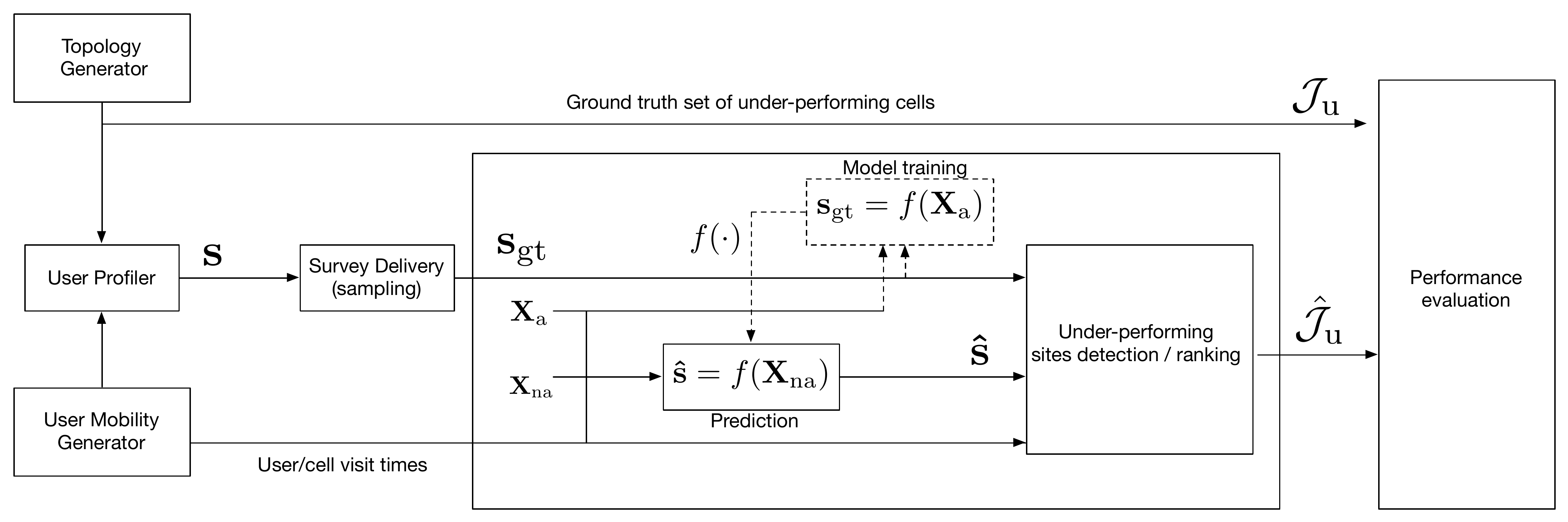}
    \caption{Architecture of a simulation framework to test the anomaly detection system.}
    \label{fig:simframe}
\end{figure}

In order to answer to questions \ref{Q1}-\ref{Q5}, we propose a simulation framework composed of several building blocks, illustrated in Figure \ref{fig:simframe}. The following Sections provide details on each component of the framework.

	\subsection{Topology Generator (TG)} The TG is responsible of generating mobile network instances composed of a set of network sites $\mathcal{J}, |\mathcal{J}|=M$, deployed in a realistic scenario (e.g., urban or rural). The TG also defines which sites $\mathcal{J}_\text{u} \subset \mathcal{J}, |\mathcal{J}_\text{u}|=\Omega<M$, are malfunctioning or under-performing in a particular network topology. The selection is performed according to a random process, specified in input. In this work we consider a uniform distribution (i.e., all network sites have the same probability of being under-performing), although an operator may use any other distribution. As an example, sites characterized by a higher level of congestion (e.g., visited by large number of users) may be selected as under-performing with a higher probability. Both the total number of sites $M$ and the number of malfunctioning sites $\Omega$ are input parameters of the simulation framework.
	
	\subsection{User Mobility Manager (UMM)}
\begin{figure}[t]
	    \centering
	    \includegraphics[scale=0.5]{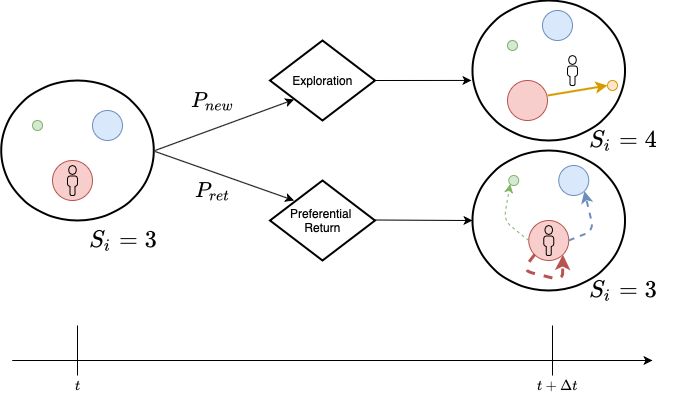}
	    \caption{Considering a generic user $i$, S equals the cardinality of the set of visited places, circles stems for the sites already visited by the user while their size represents the probability that the user visits the corresponding network site.}
	    \label{fig:mobilityalgorithm}
	\end{figure}

	The UMM models the mobility of the population of users $\mathcal{U}$ through the cellular network simulated by the TG. In particular, the UMM leverages a human mobility model which defines for the $i$-th user i) which network sites are visited and ii) for how long. Several models are available in the literature to simulate the statistical properties of human mobility \cite{hyytia2007random, lee2009slaw, munjal2011smooth, song2010modelling}. In this work we consider the model proposed in \cite{song2010modelling}, which is based on the following observations: i) humans have a periodic tendency to return to previously visited places, ii) humans spend most of their time in a few number of locations and iii) the distributions of the time spent by a user in a location $P(\Delta t)$ and the distance covered between two sightings $P(\Delta r)$ are fat-tailed, i.e. $P(\Delta r) \sim |\Delta r|^{-1-\alpha}$ and $P(\Delta t) \sim |\Delta t|^{-1-\beta}$.
	In details, the mobility model implemented in the UMM works according to different steps, as illustrated in Figure \ref{fig:mobilityalgorithm}:
	\begin{itemize}
		\item \textit{Initialization:} let $S_i$ be an integer variable which counts the number of distinct locations visited by the $i$-th user, initially set to 1. At startup, each user is associated to one site in the network topology, chosen at random.  Then, each user waits for a random period of time $\Delta t$ and eventually decides whether to explore a new location (Exploration step) or to return to an already visited site, including the current one (Preferential return step).
		\item \textit{Exploration:} with probability $P_{new} = \rho S_i{^{-\gamma}}$, the user jumps in a random direction $\theta$, uniformly distributed in the range $[0, 2\pi)$ and with a random jump length $\Delta r$. The closest site to the landing location will be visited by the user. As the user moves to this new position, the number of previously visited locations increases from $S_i$ to $S_i + 1$.
		\item \textit{Preferential Return:} with probability $1-P_{new}$, the user returns to a previously visited location with a probability proportional to the number of visits the user previously had to that location. 
	\end{itemize}
	These steps are repeated independently for each user: at each iteration the UMM updates the vector $\textbf{t}_i \in \mathbb{R}_{\ge 0}^J$ whose entries $t_{i,j}$ corresponds to the visit times of the $i$-th user in the $j$-th network site. The process ends when the total visit time for each user is equal to the simulation time horizon $T$, i.e., when $\sum_j{t_{i,j}} = T, \,\,\forall i$. The parameters controlling the user's tendency of exploring a new place $\rho$ and $\gamma$, as well as the fat-tail distribution parameters for the jump sizes $\alpha$ and the waiting times $\beta$ can be modified according to the specific case under consideration. We detail the choice of such hyper-parameters in Section \ref{sec:experiment}.
	
	\subsection{User Profiler (UP)} 
	As illustrated in Figure \ref{fig:simframe}, the UP leverages the network topology created by the TG and the mobility information output by the UMM to simulate the users (subjective) reactions $\mathbf{s}$ to the corresponding experiences in the network. As generally done in the field of QoE research \cite{orsolic2017machine, zhang2018distributed, pimpinella2019towards}, in this work we assume the user reactions to be binary, i.e., $\mathbf{s} \in [0,1]^N$. In details the $i$-th user reaction $s_i$ is defined as:
    \begin{equation}
       s_i=
       \begin{cases}
         1, & \text{if the $i$-th user is dissatisfied with her network service}\\
         0, & \text{otherwise}.
       \end{cases}
     \end{equation}
	 
It is well known from the literature that the duration of a network disservice has great impact on the experience perceived by a user. As an example, in the case of video streaming, QoE is primarily influenced by the frequency and duration of stalling events \cite{hossfeld2011quantification, nam2016qoe}. Similarly, for web browsing, the number and duration of IRAT handovers is shown to have a strong negative impact on user experience \cite{balachandran2014modeling}. In both cases users are observed to tolerate a certain amount of disservice before expressing a negative opinion: for video streaming, one stalling event per clip is acceptable as long as its duration is below 3 seconds, while for web browsing a single IRAT handover is generally tolerated.
	
Following these observations, it is reasonable to link the user satisfaction $s_i$ to the time spent in under-performing or malfunctioning network sites. Operatively, the UP leverages the set of under-performing sites $\mathcal{J}_u$ and the user visit times $t_{i,j}$ to generate user satisfaction according to the following:
    \begin{equation}
       s_i=
       \begin{cases}
         1, & \text{if}\,\, \sum_{j \in \mathcal{J}_c} t_{i,j} \geq u_i T\\
         0, & \text{otherwise}.
       \end{cases}
     \end{equation}

where $T$ is the simulation time horizon and $u_i$ is a percentage value corresponding to the \textit{user tolerance}. In other words, we assume that each user has a specific patience level with respect to negative network experiences. Intuitively, the higher the tolerance of a user the more she will tolerate low service quality during her network activity. To model the heterogeneity of the users, we assume that the user tolerance $u_i$ is a Gaussian-distributed random variable with mean $\mu$ and standard deviation $\sigma$, i.e., $u_i \sim \mathcal{N}(\mu, \sigma^2)$. Later in Section \ref{sec:sensan} we will discuss about the choice of their values.

Finally, we observe that the reported user satisfaction depends also on factors completely unrelated with the network service itself, such as the ones relative to users personal attitudes and expectations~\cite{boz2019mobile}. The UP models such noisy behaviours by generating a percentage $\psi$ of the satisfaction labels $s_i$ at random, regardless of the sites visited by users and their tolerance. Again, $\psi$ represents a hyper-parameter of the model that can be set by the operator to simulate different population types.	
	
	\subsection{Survey Delivery (SD)} 
\begin{figure}[t]
\centering     
\subfigure[RD: low network coverage \newline scenario]{\label{fig:random}\includegraphics[width=0.38\textwidth]{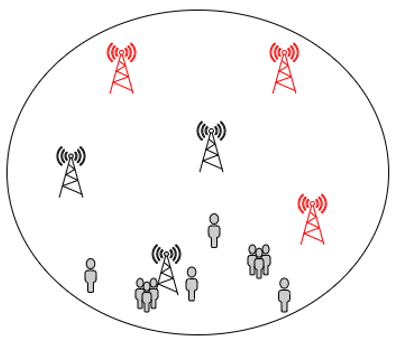}}
\subfigure[OD: high network coverage \newline scenario]{\label{fig:optimum}\includegraphics[width=0.4\textwidth]{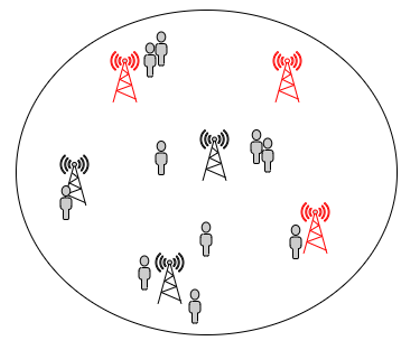}}
\caption{Strategies for the delivery of satisfaction surveys: a network coverage perspective. User icons refer to the users who replied to the received surveys, while blackish and redish base stations refer to regular and anomalous network sites respectively.}
\label{fig:deliveries}
\end{figure}

	Any crowdsourcing-based network monitoring system is limited by the associated \textit{network coverage}, i.e., the percentage of sites visited by users answering the surveys, as it is not possible to detect under-performing sites for which no information from users is available.
	As aforementioned, the number of users which answer to surveys is generally small compared to the total number of customers, and the set of corresponding \textit{Ground Truth} responses $\mathbf{s}\textsubscript{gt}$ is much smaller than $\mathbf{s}$. The SD component of the proposed framework simulates the process of administrating satisfaction surveys to the customers, and can therefore be tought of as a sampling process of the true user reactions.
In this paper we consider two scenarios for the surveys delivery strategy:
	
\begin{itemize}
		\item Random Delivery (RD): in the most general case, we assume that the set of users answering the surveys is randomly sampled from the total population. This strategy is represented in Figure \ref{fig:random}, where user icons represent those users who replied to the received survey. In this (unlucky) example the operator has low network coverage, as the received responses do not cover under-performing sites in the network.
		\item Optimized Delivery (OD): with this policy, depicted in Figure \ref{fig:optimum}, the operator delivers surveys in a way to maximize the network coverage. This is done by leveraging the user-visit times $t_{i,j}$ to set up an optimization problem (formalized in Section \ref{sec:ILP}) that selects the smallest set of users whose responses would allow to maximize the number of visited sites. Clearly, in this case we assume that the operator puts in place an incentive strategy for the voting procedure, such that a user selected from the optimization problem is rewarded for its answer (e.g. with premium data plan access for limited time or other incentives). 		

\end{itemize}

Regardless of the chosen delivery strategy, the SD block samples the set of users reactions $\mathbf{s}$ and returns a set of GT users feedbacks $\mathbf{s}\textsubscript{gt}$, which is input to the under-performing sites ranking and detection algorithm.

\subsection{Under-Performing Sites Ranking and Detection Algorithm}\label{sec:satprediction}
At this point, the detection system mentioned at the end of Section \ref{sec:system} can be used to detect/rank under-performing sites in the network. The ranking algorithm will leverage: i) the user-specific cell visit times information generated by the UMM; ii) the set of GT survey responses generated by the SD block and a pre-trained ML model $f(\cdot)$, to predict the satisfaction feedback of all those users who did not answer a survey. Note that the ML model is assumed to be already trained and detailing which learning features belong to $\mathbf{X}_\text{a}$ and $\mathbf{X}_\text{na}$ is outside the scope of this paper. For the sake of clarity, we underline that the ranking algorithm is blinded about the true location of the under-performing sites, i.e., it does not know which network site belongs to $\mathcal{J}_u$.

For each network site in the network topology, the algorithm computes a score $r_j$ according to the following procedure: 
\begin{enumerate}
	\item First, the set $\mathcal{V}_j$ of all the dissatisfied visitors of site $j$ is selected. Note that $\mathcal{V}_j$ contains all those users such that $t_{i,j}>0$ and the associated ground truth or predicted satisfaction is 0. We recall that a user visits multiple sites and its dissatisfaction may be due only to one of them. To cope with this, we tighten the time constraint as it follows:
	\begin{equation}
		t_{i,j} > \xi \sum_j t_{i,j}
	\end{equation}
where $\xi$ is a percentage that acts as an activation threshold for considering site $j$ as responsible to the experience of the $i$-th user. Further details about the choice of the value of $\xi$ will be given in Section \ref{sec:sensan}. 
	
\item Then, the site score is computed by summing all normalized user visit times above the threshold $\xi$, that is:
	\begin{equation}
    r_j = \sum_{i \in \mathcal{V}_j} \frac{t_{i,j}}{\sum_j t_{i,j}}.
	\end{equation}
\end{enumerate}

Finally, network sites are ranked in descending order according to $r_j$ and the operator may use such an information to prioritise upgrading investments in the network. Here we assume that the operator has a budget for upgrading $k$ network sites. Feeding the value of $k$ into the detection system allows to output a set $\hat{\mathcal{J}}_\text{u}$, $|\hat{\mathcal{J}}_\text{u}| = k$, containing the first $k$ sites of the ranked list.

In the following Section we run the simulation framework in different scenarios and we compare the ranked set $\hat{\mathcal{J}}_\text{u}$ with the true set of under-perfoming cells $\mathcal{J}_\text{u}$ to assess the detection performance.

\section{System evaluation}\label{sec:experiment}
We use the proposed simulation framework to perform several experiments, with the goal of answering questions (\ref{Q1}-\ref{Q5}). This section is organized as follows: first, we provide details on the experimental setup in Section \ref{sec:overview}. Then, Section \ref{sec:sensan} focuses on the relationship between users satisfaction profile and the detection performance, providing an answer to Questions \ref{Q1},\ref{Q2}. Finally, Section \ref{sec:tradeoff} comments on the impact that both the surveys delivery strategy and the satisfaction prediction errors have on the overall ranking task, thus answering Question \ref{Q3}, \ref{Q4} and \ref{Q5}.   

\subsection{Experiments Overview}\label{sec:overview}
\begin{figure}[t]
    \centering
    \includegraphics[width=0.5\textwidth]{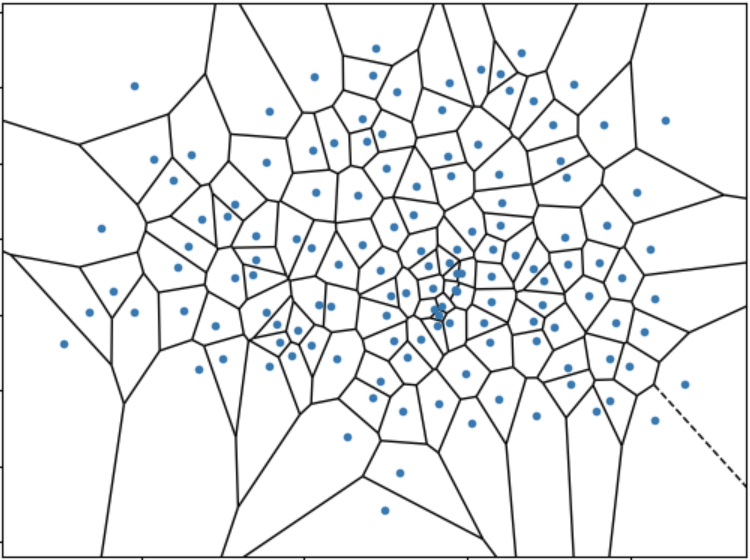}
    \caption{Voronoi representation of the considered network.}
    \label{fig:voronoimap}
\end{figure}
We feed the Topology Generator with information gathered from a a real cellular network currently operative in a middle-sized European city. The cellular network is composed of $136$ network sites deployed in an area of approximately 180Km$^2$, whose locations are illustrated in Figure \ref{fig:voronoimap}. 
We consider three different densities of users per network site, corresponding to population sizes equal to $100$k, $10$k and $1$k users. 
Moreover, regardless of the population size, we consider two different mobility scenarios according to the value of the hyper-parameter $\gamma$ that is input to the UMM:
\begin{itemize}
    \item \textit{Scenario 1 (S1)}: this case reproduces the setup described in \cite{song2010modelling}, where a dataset containing one-year period trajectories of three million anonymized mobile-phone users is used to statistically estimate the values of the hyper-parameters. In this case, $\gamma$ is set equal to $0.21$; 
    \item \textit{Scenario 2 (S2)}: we reproduce the users mobility patterns observed in a dataset of $1500$ anonymised customers in the same cellular network used to feed the TG for a period of 1 month. In this case, $\gamma$ is set equal to $3$. Therefore, this scenario is characterized by a lower tendency of the users to visit new sites compared to the first scenario.
\end{itemize}
For what regards the others parameters input in the UMM (i.e., $\alpha$, $\beta$,  and $\rho$), they are set to values estimated in \cite{song2010modelling} for both scenarios, that is $\alpha = 0.55, \beta = 0.8$ and $\rho = 0.6$.

\begin{figure}[t]
\centering
  \subfigure[Scenario S1.]{\label{fig:dist_TM} 
   \includegraphics[width=0.47\linewidth]{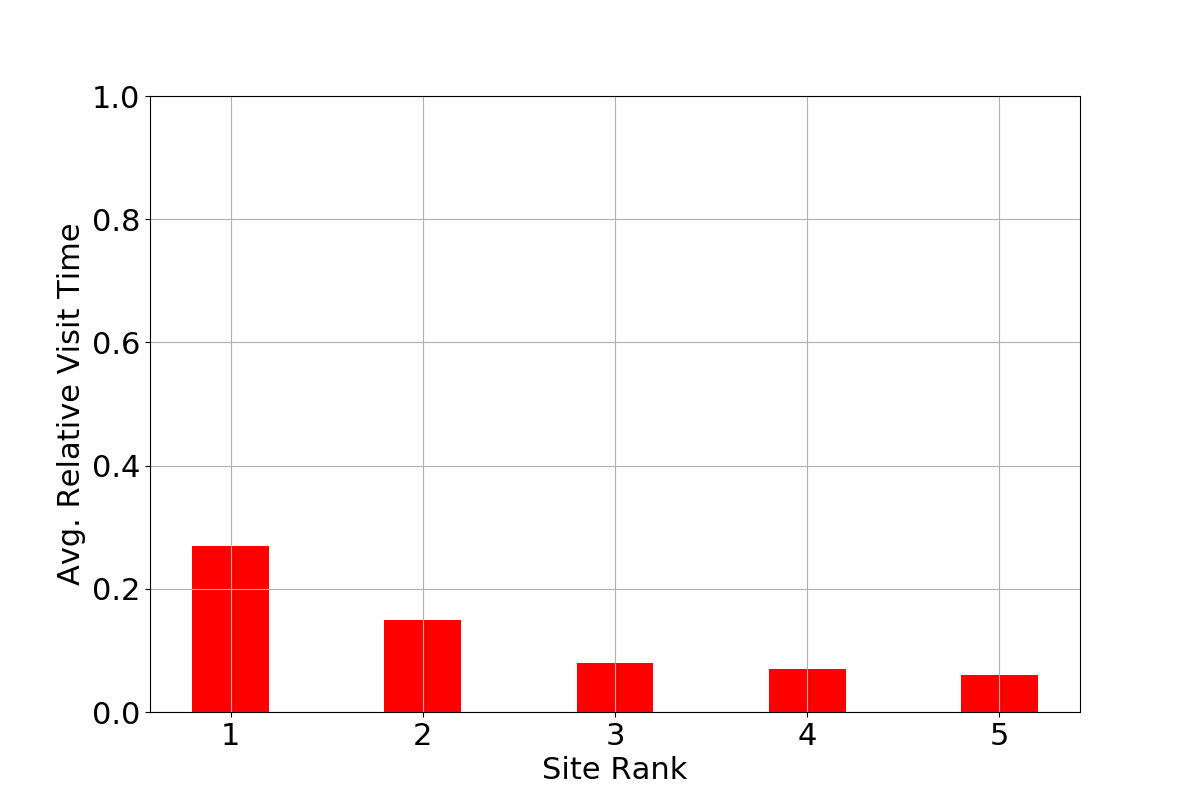}}
\subfigure[Scenario S2 and empirical data.]{\label{fig:dist_EM}
    \includegraphics[width=0.47\linewidth]{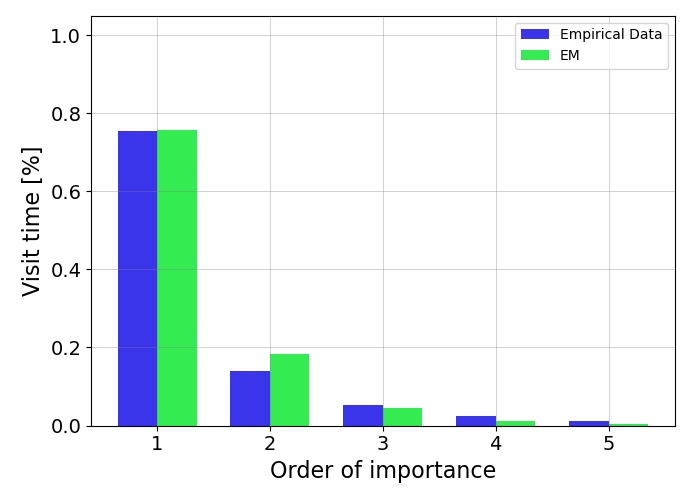}}
\caption{Average distribution of users visit time versus the site's Order of Importance, for scenarios S1 (red bins) and S2  (blue bins), and empirical mobility data (green bins).}
\label{fig:scenarios}
\end{figure}

We plot in Figures \ref{fig:dist_TM} and \ref{fig:dist_EM} the average proportion of visit time resulting from the UMM for the case of $100$k users for the two scenarios, ranked in decreasing order. The first bar refers to the average proportion of time spent by users in the most visited site, the second bar refers to the second most visited site and so on. As one can see, in both scenarios the distributions have a negative exponential trend, with the five most visited sites representing on average more than 60\% and 95\% of the overall users visit time in the network for S1 and S2 scenarios, respectively. 
\par For what concerns the generation of the under-performing sites and the users profiling, we set the values of $\Omega$ (i.e., the number of under-performing sites in the network) to $\lfloor 0.1M \rfloor$, while $\mu$ and $\sigma$ (i.e., mean and variance of the random variable $u_i$ that controls users tolerance) are set so that the percentage of dissatisfied users ranged between $15\%$ and $30\%$ of the whole population. This is because cellular users feedbacks are typically unbalanced ~\cite{huang2015telco, pimpinella2019towards, boz2019mobile}, i.e., the class of satisfied users is usually much larger then the class of dissatisfied ones. 
Finally, we leave to the next Section the discussion about the choice of the value of $\xi$, where we will also comment its relationship with the profile of the visiting population (i.e., with the hyper-parameters $\mu$ and $\sigma$). 
\subsection{Detection Performance and Users Heterogeneity} \label{sec:sensan}
As a first experiment, we use the simulation framework to find answers to questions \ref{Q1} and \ref{Q2}. We leave aside the problem of predicting user satisfication, deactivating the sampling process in the survey delivery block and assuming an ideal scenario in which the operator has knowledge of to the true satisfaction $\mathbf{s}$ for all users. At the same time, we are intersted in understanding how users heterogeneity impacts on the process of detecting under-performing sites. Therefore, we analyse the impact of parameters $\xi$ and $\mu$ on the detection performance. In particular, $\xi$ is varied between $0.1$ and $0.5$ while $\mu$ takes values in [$0.05$, $0.15$, $0.25$, $0.35$]. Note that for each value of the average user tolerance $\mu$, the corresponding value of $\sigma$ is adjusted in order to let the fraction of dissatisfied users be within $15\%$ and $30\%$.

We recall that the framework outputs a set $\hat{\mathcal{J}}_u$ containing the $k$-th, where $k$ is an input parameters which depends on the operator financial budget. The performance metrics used for evaluating the detection performance are the \textit{Precision} and \textit{Recall} \textit{at} $k$ ($P@k, R@k$), defined as:
\begin{align}
&P@k = \frac{\vert \hat{\mathcal{J}}_u(k)  \cap  \mathcal{J}_u \vert}{\vert \hat{\mathcal{J}}_u(k) \vert} \\
&R@k = \frac{\vert \hat{\mathcal{J}}_u(k)  \cap  \mathcal{J}_u \vert}{\vert \mathcal{J}_u \vert} 
\end{align}
where the numerators correspond to the number of correctly detected sites, while the denominators equal $k$ and $\Omega$ respectively.

As one can see, $P@k$ is defined as the proportion of the top-$k$ ranked network sites that are actually under-performing. On the other hand, $R@k$ corresponds to the proportion of correctly detected under-performing sites.
We perform several experiments with different values of $\xi$ and $\mu$. Since an operator is unaware of the true number of under-performing sites, we evaluate the performance for different values of $k = 1 \ldots M$, evaluating each time the measures $P@k$ and $R@k$. Finally, for a fixed couple ($\xi$, $\mu$) we first compute the Precision-Recall ROC curve at different values of $k$, and then we summarize the performance of the system with the \textit{Area Under the Curve} value, AUC($\xi, \mu$). We highlight that the AUC summarizes the detection performance for all possible values of $k$.

\begin{figure}[t]
\centering     
\subfigure[Scenario 1]{\label{fig:caseTM}\includegraphics[width=0.45\textwidth]{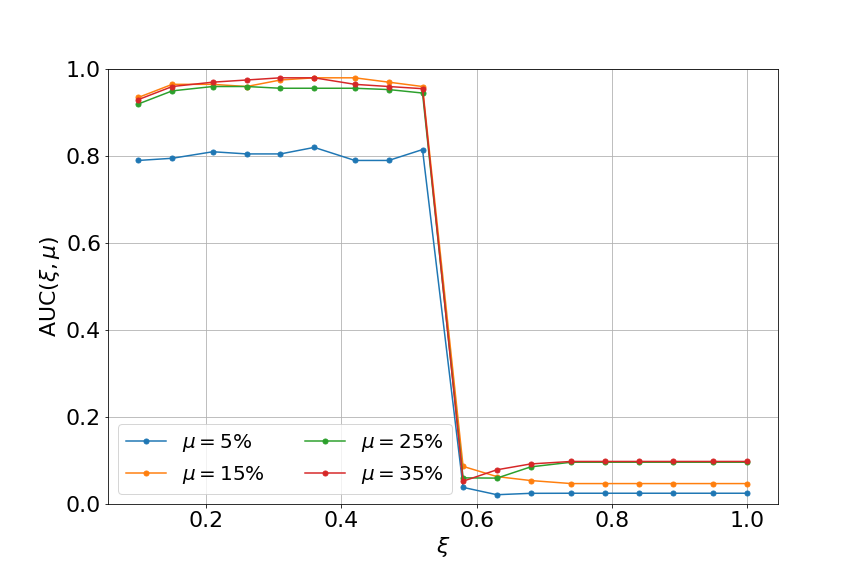}}
\subfigure[Scenario 2]{\label{fig:caseEM}\includegraphics[width=0.45\textwidth]{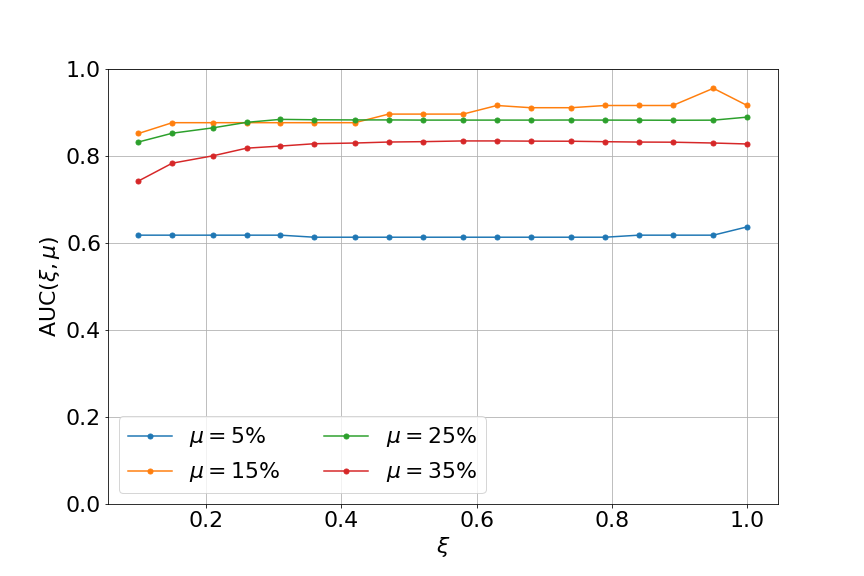}}
\caption{AUC vs $\xi$ for different average users tolerance to bad network events $\mu$. The population size equals $100k$ users, who move according to mobility scenarios S1 (left) and S2 (right).}
\label{fig:RvsTTAvsUT}
\end{figure}

We run the tests $10$ times, each time generating a new random set of under-performing network sites. 
Figures \ref{fig:caseTM} and \ref{fig:caseEM} plot the average AUC values of the detection process when applied to S1 and S2 scenarios, respectively. In both scenarios, we observe that when the satisfaction feedbacks are retrieved from a population of excessively \textit{touchy} users (i.e., when $\mu=0.05$), the system AUC lowers on average by $15\%$ with respect to the other population profiles. Referring to S1 (Figure \ref{fig:caseTM}), we observe that the detection system performs similarly for each $\mu$ greater than $5\%$. Differently, in scenario S2 (Figure \ref{fig:caseEM}), the AUCs are similar for $\mu$ equal to $0.15$ and $0.25$ while it is on average $8\%$ lower for $\mu = 0.35$. 

From these observations we conclude that:
\begin{enumerate}
\item The detection performance depends i) on the way users move throughout the network and ii) on their subjective profiles. In general, considering that the true value of $\mu$ is unkwown and uncontrollable by the operator, the ranking algorithm yields good detection performance in both scenarios, with AUC greater than $80\%$ and $60\%$ for S1 and S2 respectively. This answers to question \ref{Q1} introduced in Section \ref{sec:system}.
\item Regardless of the average tolerance $\mu$ of the population, we observe that the detection performance are stable with respect to the value of $\xi$, highlighting a certain inherent robustness of the system. This is promising, as an operator doesn't need to worry about i) estimating $\mu$ or ii) tuning $\xi$ with excessive care. As a rule of thumb, setting $\xi$ as the mean of the average times spent in the most visited and second most visited site provides a good working point. This provides the answer to  question \ref{Q2}.
\end{enumerate}

\subsection{Satisfaction Prediction Trade-Off} \label{sec:tradeoff}
To tackle question \ref{Q3}, we analyze the performance of the system in a more realistic case where the operator has only partial information about users satisfaction feedbacks. In particular, we investigate whether it is more reliable for an operator to perform detection considering only GT users feedbacks (i.e., only $\mathbf{s}\textsubscript{gt}$) or it is convenient to include also the predicted satisfaction labels (i.e., also $\hat{\mathbf{s}}$). 
Without loss of generality, the analysis is performed fixing the values of $\mu$, $\sigma$ and $\xi$ to $0.25$, $0.029$ and $0.2$, respectively. 
The performance metric used for this analysis is the Recall at $\Omega$, $R@\Omega$. Note that $R@\Omega$ has a maximum value of 1, if all under-performing cells are detected.
The size of the GT users sets are fixed to $1\%$ of the overall population size, i.e., we will assume a users response rate to satisfaction surveys of $1\%$, as observed in ~\cite{pimpinella2019towards}. Consequently, the three populations of $1$k, $10$k and $100$k users will be respectively characterized by an average density of $0.073$, $0.73$ and $7.35$ GT users per network site, which we refer to as Low, Medium and High density. 
Concerning the delivery strategies, the OD optimization problem is solved by setting $n=3$. For a fair comparison, the OD budget $B$ is equal to the number of GT surveys used in the RD case.


\subsubsection{Users QoE Prediction for Anomaly Detection} \label{sec:perfcloud}

\begin{figure}[!t]
\centering
\subfigure[RD strategy, $100k$ users]{\label{fig:RD 100k TM} 
   \includegraphics[width=0.9\columnwidth]{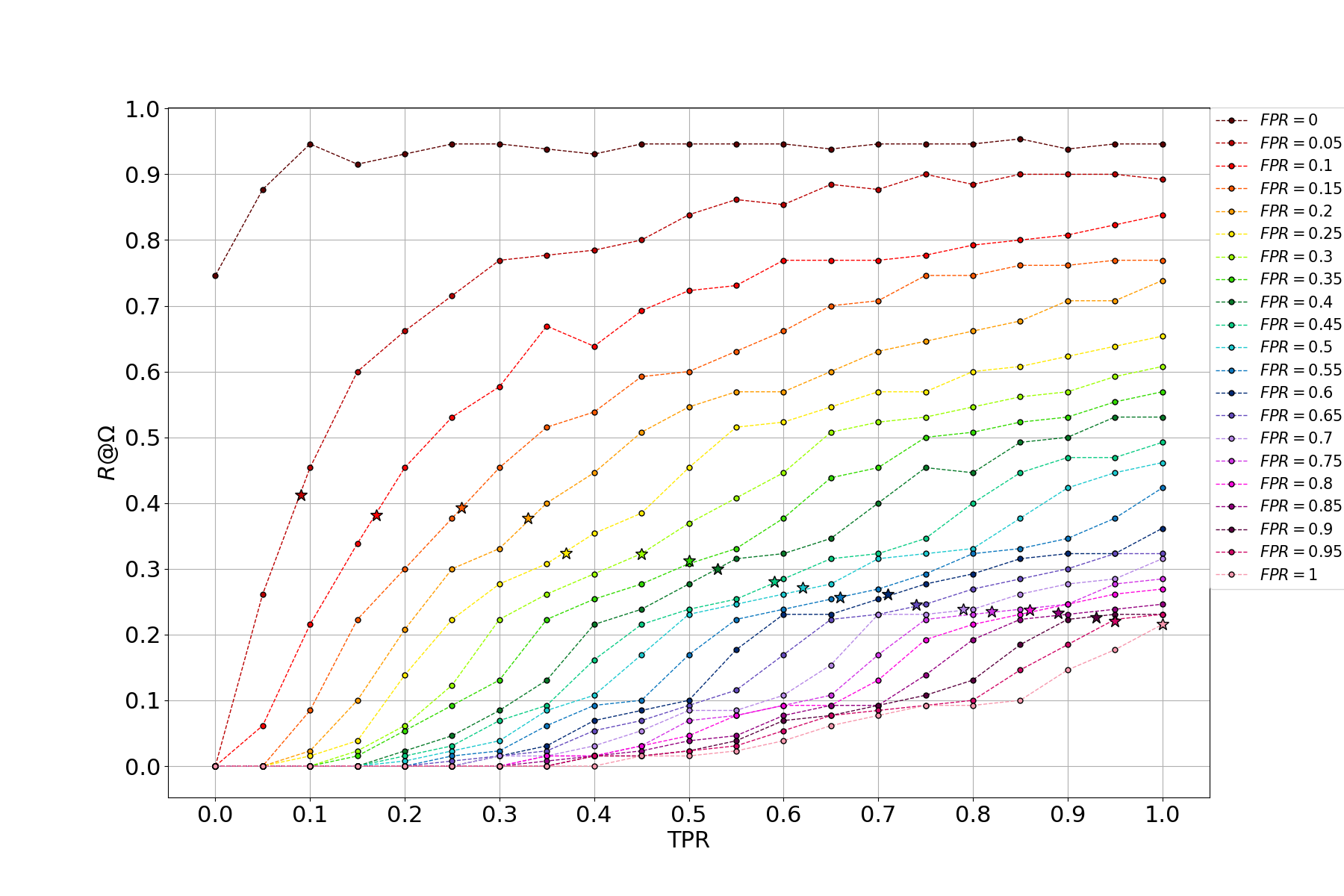}}
\subfigure[OD strategy, $100k$ users]{\label{fig:OD 100k TM}
    \includegraphics[width=0.9\columnwidth]{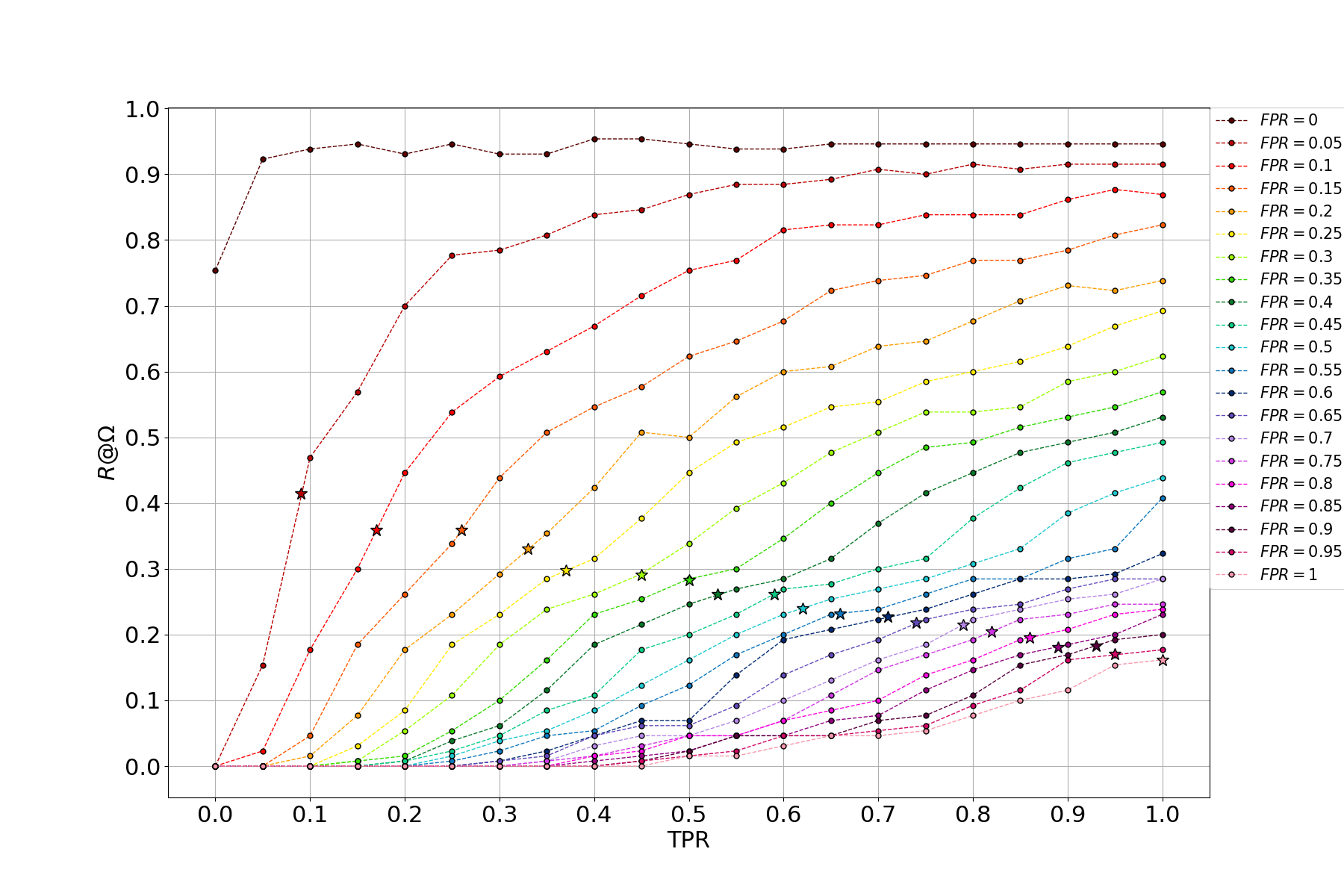}}
\caption{Performance Clouds for a population of 100k Users moving according to S1.}\label{fig:TM_pc_100k}
 \end{figure} 
 \begin{figure}[!t]
\centering
\subfigure[RD strategy, $100k$ users]{\label{fig:100k RD EM}
    \includegraphics[width=0.9\columnwidth]{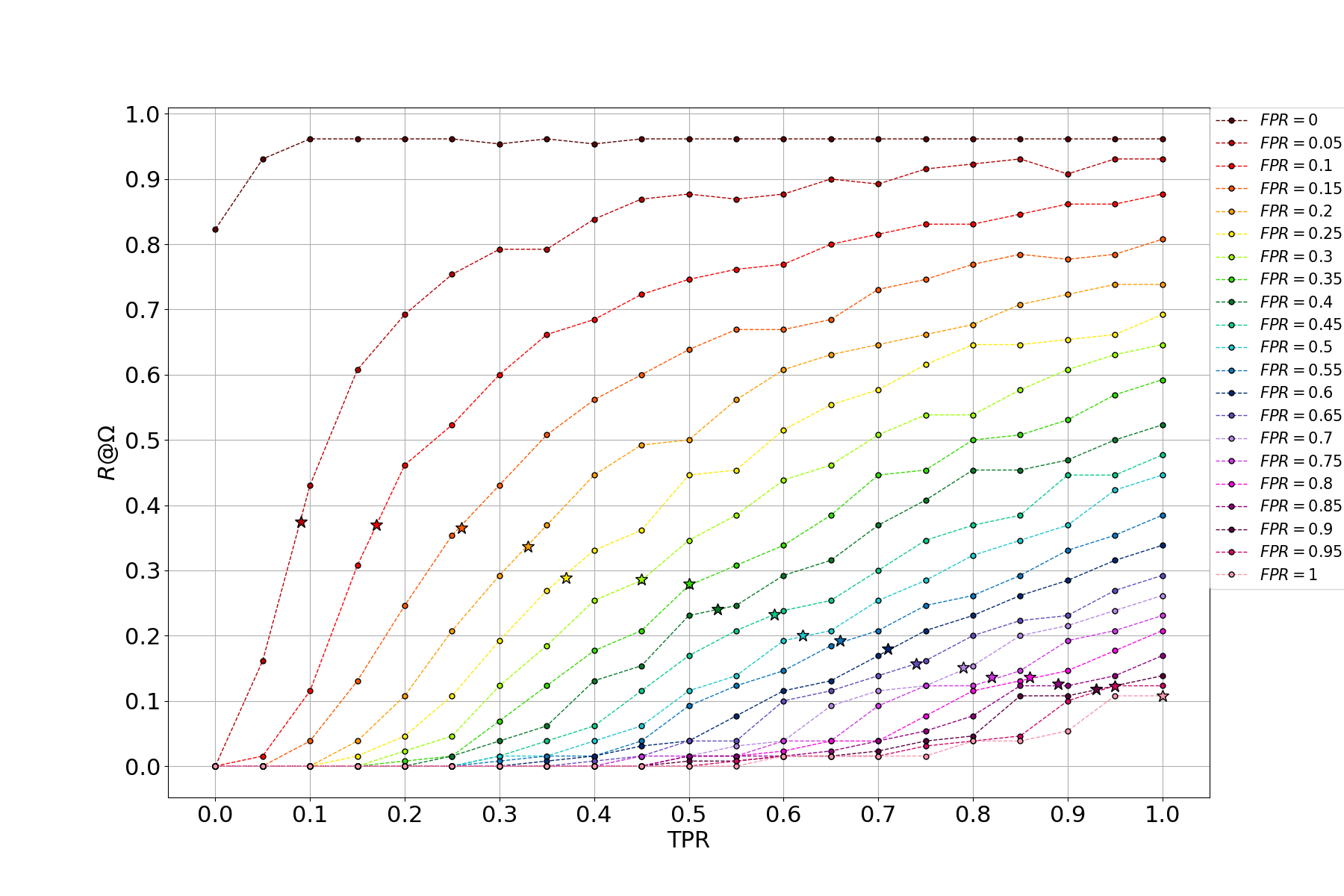}}
\subfigure[OD strategy, $100k$ users]{\label{fig:100k OD EM}
   \includegraphics[width=0.9\columnwidth]{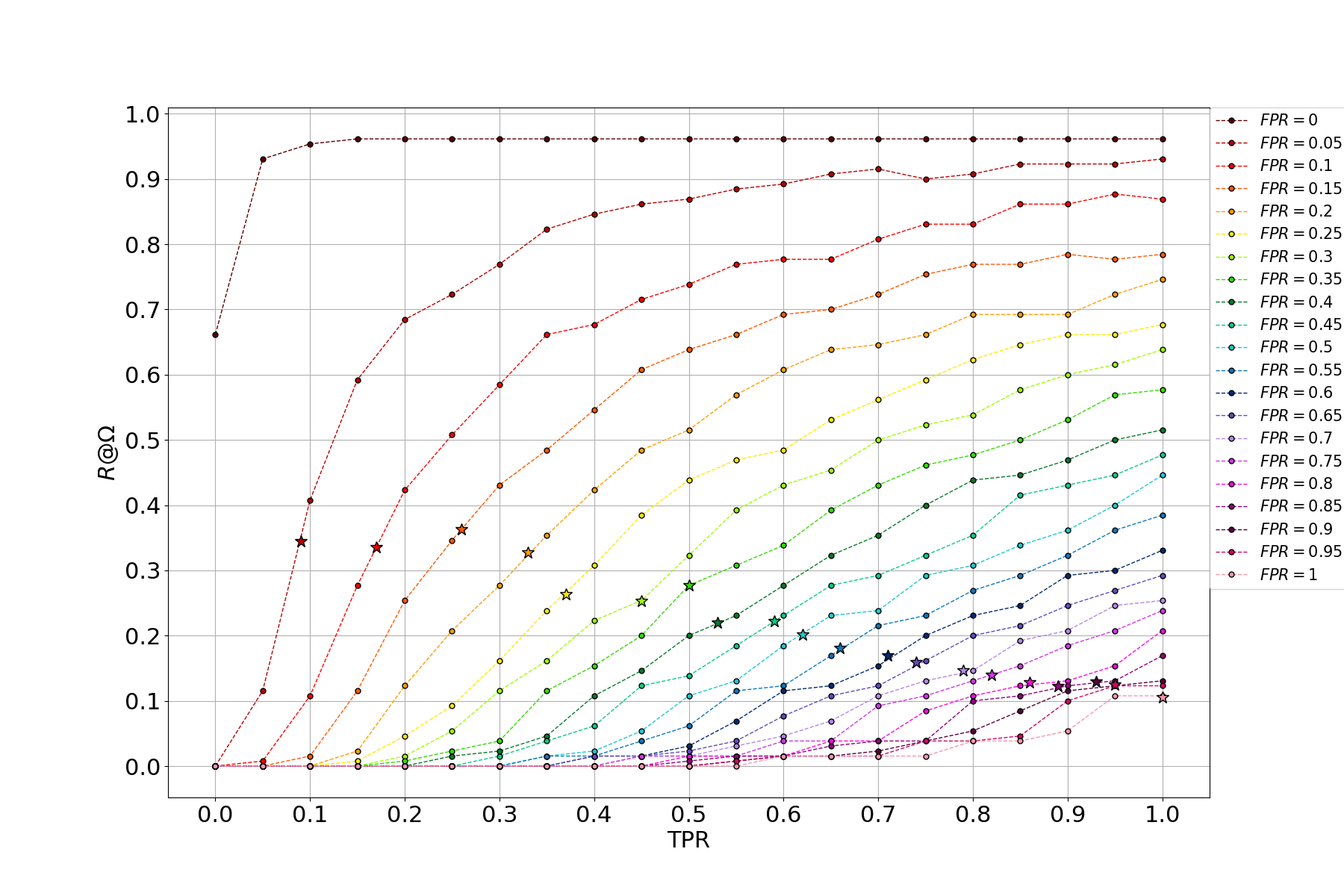}}
\caption{Performance Clouds for a population of 100k Users moving according to S2.}\label{fig:EM_pc_100k}
\end{figure}

Any machine learning algorithm an operator can use to predict the user satisfaction $\hat{\mathbf{s}}$ will be characterized by a certain prediction error. Considering the nature of the satisfaction prediction problem, we assume the availability of a binary classifier and express its performance with the False Positive Rate (FPR) and the True Positive Rate (TPR) metrics. In details, the FPR corresponds to the rate of false alarms (satisfied users predicted as dissatisfied), while the TPR corresponds to the recall of the classifier (percentage of dissatisfied users detected). We observe that ML classifiers are characterized by several FPR and TPR working points, which can be traded-off by tuning a decision threshold. To perform a comprehensive analysis, we run the simulation framework assuming the availability of several ML classifiers in order to cover all possible FPR and TPR working points. In particular, we let both the FPR and the TPR vary between $0$ and $1$ with step-size equal to $0.05$, thus analyzing $400$ different performance points.

As an example, Figure \ref{fig:TM_pc_100k} shows the obtained $R@\Omega$ for the case of 100k users moving according to Scenario 1, where 1k user satisfaction grades are sampled according to the RD strategy and the remaining are predicted with a ML classifier.
Curves with different colours refer to classifiers with different (and fixed) FPR values, while the TPR is shown on the abscissa. Fixing a value of FPR (i.e., referring to one of such curves), allows to observe the recall of the sites detection system versus the TPR of the user satisfaction classifier. The colored stars refer to the performance of the classifier proposed in \cite{pimpinella2019towards}.

Considering a population of $100k$ users (Figures \ref{fig:TM_pc_100k} and \ref{fig:EM_pc_100k}) we observe that: i) for a fixed TPR, the detection accuracy increases with decreasing FPR values; ii) for a fixed FPR, the detection accuracy improves with increasing TPR values; iii) for a fixed value $\Delta$, decreasing the FPR by $\Delta$ is more beneficial than increasing the TPR by the same value.
These observations suggest that i) predicting that a satisfied user is dissatisfied (i.e., having a false positive) is more detrimental for the detection process than missing a dissatisfied user (i.e., missing a true positive) and ii) when deciding the FPR/TPR tradeoff of its classifier, an operator should prefer working points at low FPR rather than at high TPR. Moreover, this holds regardless of the population size, the mobility type and the surveys delivery strategy, which are illustrated in Figures \ref{fig:TM_pc_others} and \ref{fig:EM_pc_others}. 

This provides an answer to question \ref{Q3}.


\subsubsection{To Predict or not to Predict?} \label{tab:GTvsClass}
\begin{table}[t]
\centering
\caption{S1: Working points of a real binary classifier that yield best anomaly detection accuracy.}
\label{tab:BA_S1}
\begin{tabular}{|c|c|c|c|c|}
\hline
\textbf{GT Users/Site} & \textbf{Delivery} & (FPR\textsubscript{C},  TPR\textsubscript{C})(\%) & $R\textsubscript{C}@\Omega$ (\%) & $R\textsubscript{gt}@\Omega$ (\%) \\ \hline

\multirow{2}{*}{Low}                 & RD                             & (20,33)                                                                             & 28     & 7                \\ \cline{2-5} 
                                    & OD                             & (35,50)                                                                             & \textbf{29}     & 9                \\ \hline

\multirow{2}{*}{Medium}                & RD                             & (15,26)                                                                             & 39              & 39               \\ \cline{2-5} 
                                    & OD                             & (5,9)                                                                               & 41              & \textbf{53}      \\ \hline
                                    
\multirow{2}{*}{High}               & RD                             & (5,9)                                                                               & 42              & {75}      \\ \cline{2-5} 
                                    & OD                             & (5,9)                                                                               & 42              & \textbf{82}      \\ \hline                                 

\end{tabular}
\end{table}
\begin{table}[t]
\centering
\caption{S2: Working points of a real binary classifier that yield best anomaly detection accuracy.}
\label{tab:BA_S2}
\begin{tabular}{|c|c|c|c|c|}
\hline
\textbf{GT Users/Site} & \textbf{Delivery} & (FPR\textsubscript{C}, TPR\textsubscript{C}) (\%) & $R\textsubscript{C}@\Omega$ (\%) & $R\textsubscript{gt}@\Omega$ (\%) \\ \hline
\multirow{2}{*}{Low}                 & RD                             & (15,26)                                                                             & 22     & 6                \\ \cline{2-5} 
                                    & OD                             & (20,33)                                                                             & \textbf{25}     & 8                \\ \hline

\multirow{2}{*}{Medium}                & RD                             & (15,26)                                                                             & 35              & 50      \\ \cline{2-5} 
                                    & OD                             & (5,9)                                                                               & 39             & \textbf{58}      \\ \hline

\multirow{2}{*}{High}               & RD                             & (10,16)                                                                             & 39              & 81      \\ \cline{2-5} 
                                    & OD                             & (15,26)                                                                             & 39              & \textbf{84}      \\ \hline

\end{tabular}
\end{table}

\begin{figure}[t]  
\centering
\subfigure[RD strategy, $10k$ users]{\label{fig:RD 10k TM} 
   \includegraphics[width=0.45\linewidth]{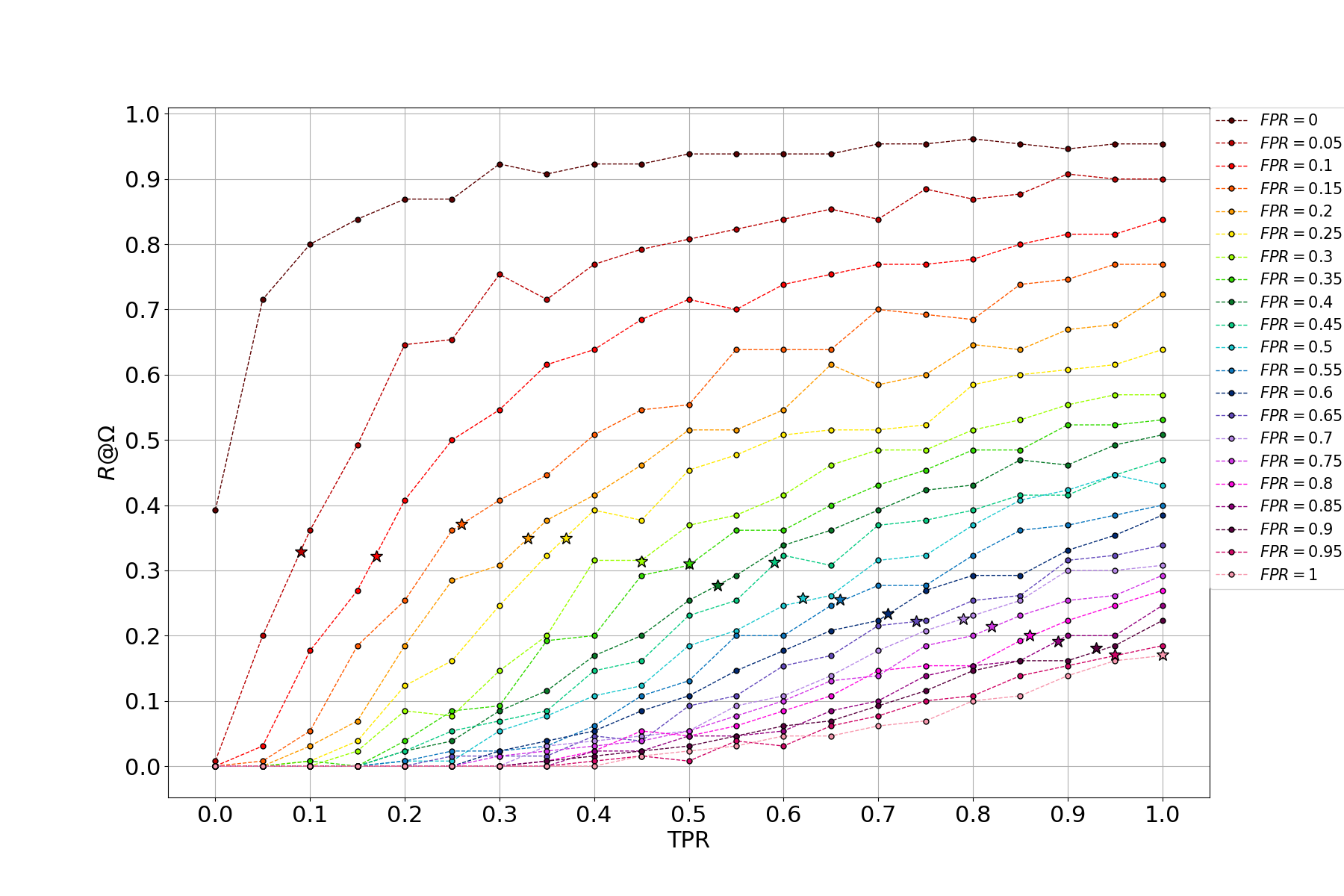}}
\subfigure[OD strategy, $10k$ users]{\label{fig:OD 10k TM}
    \includegraphics[width=0.45\linewidth]{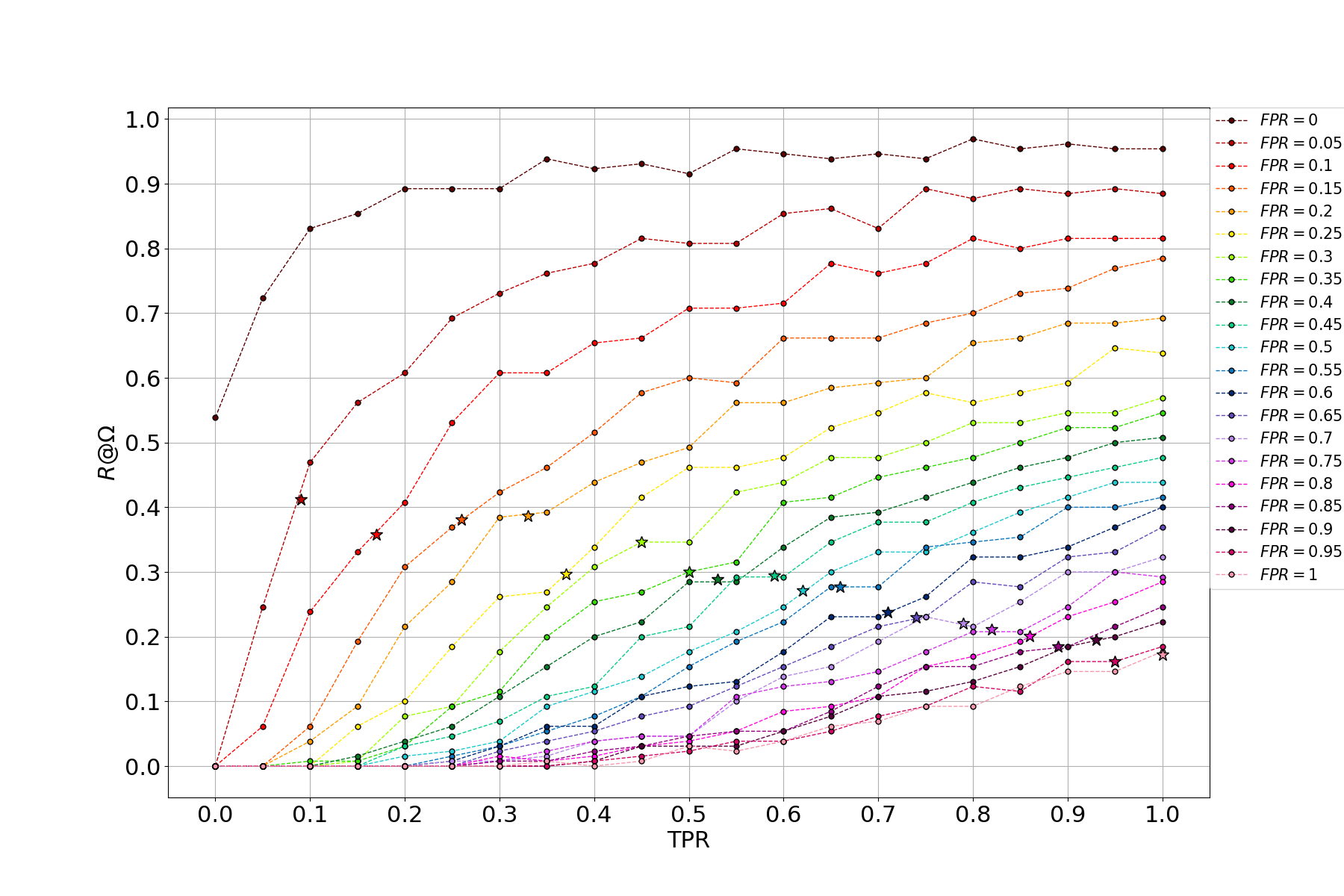}}
 \subfigure[RD strategy, $1k$ users]{\label{fig:RD 1k TM} 
   \includegraphics[width=0.45\linewidth]{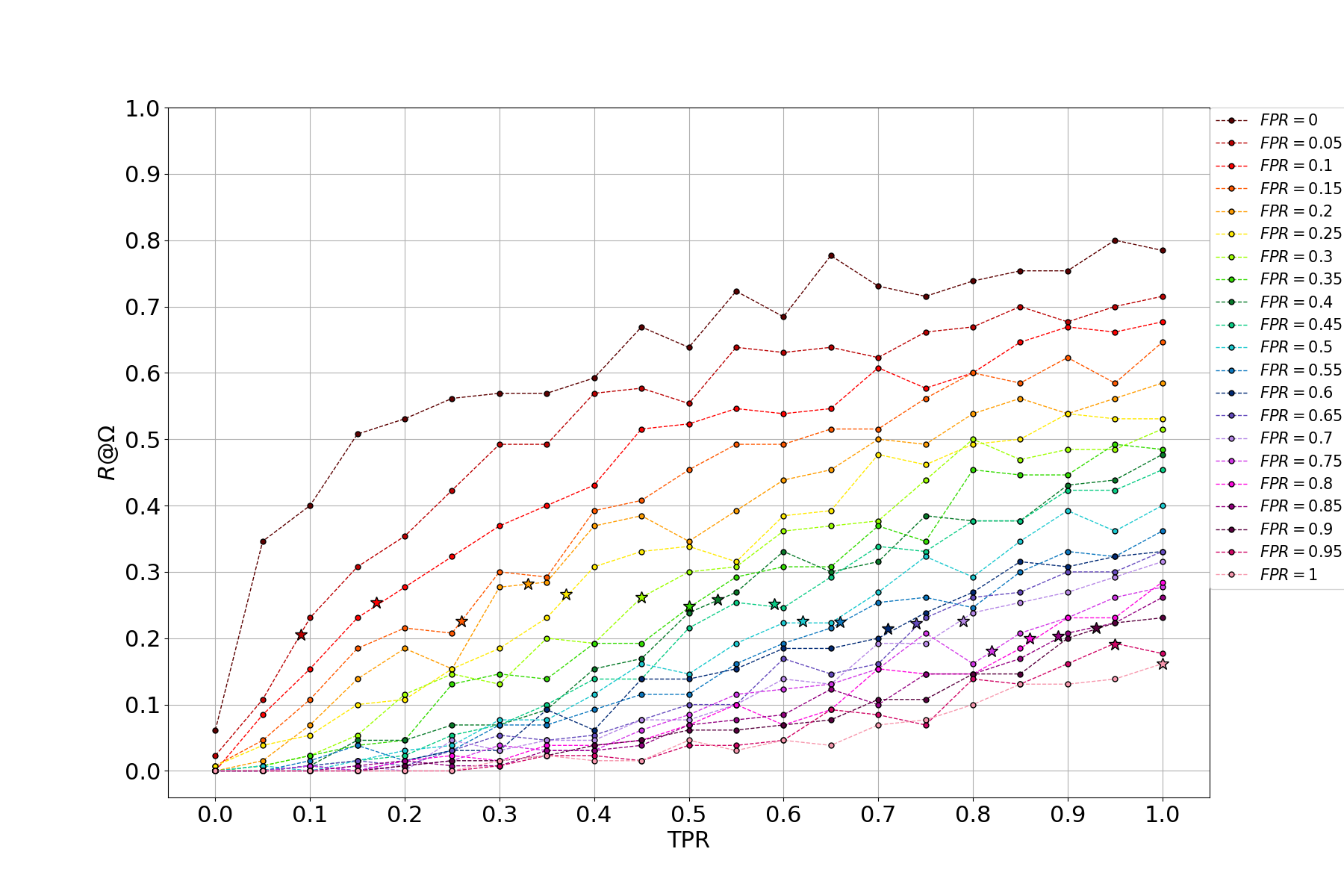}}
\subfigure[OD strategy, $1k$ users]{\label{fig:OD 1k TM}
    \includegraphics[width=0.45\linewidth]{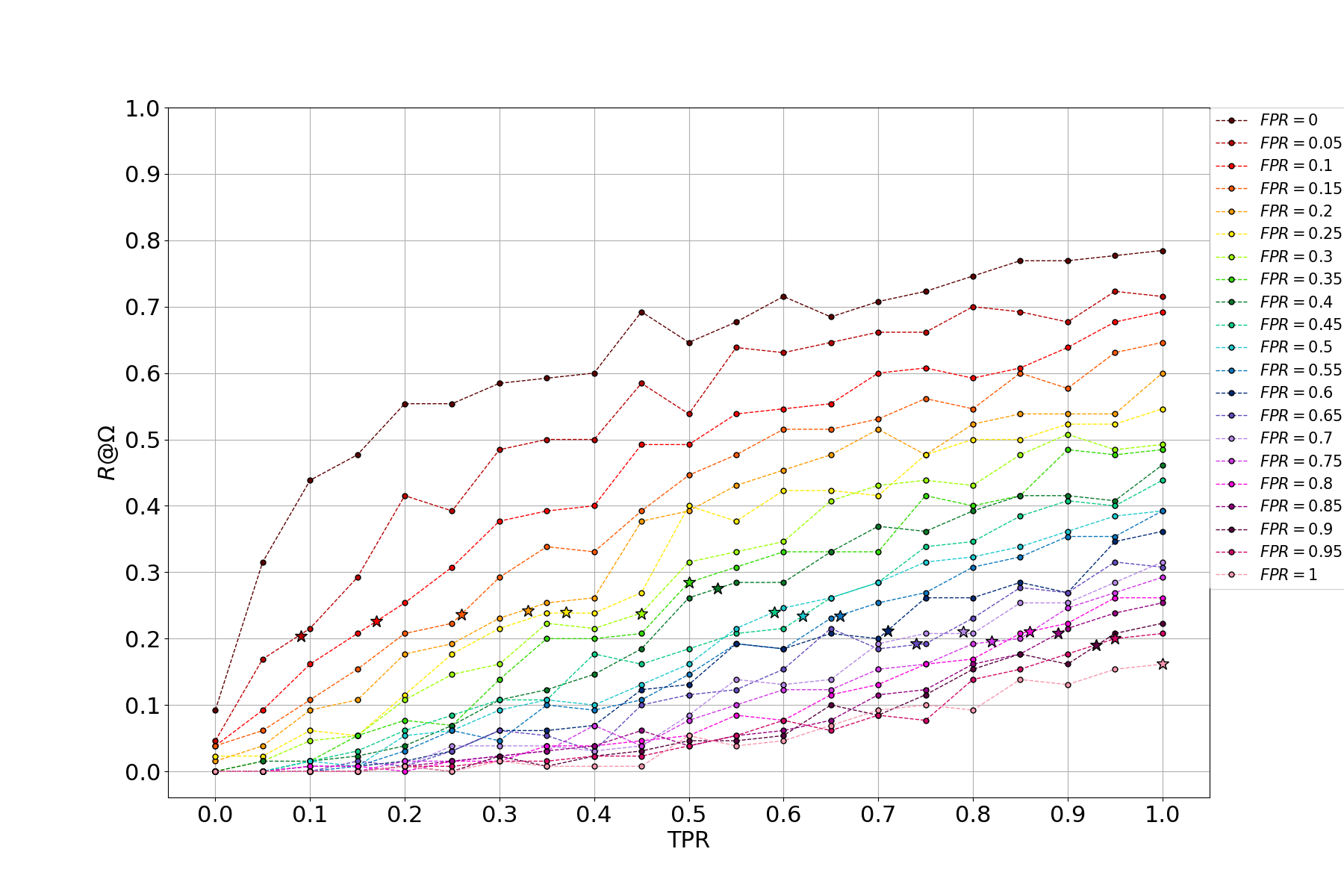}}
\caption{Scenario S1, 10k and 1k Users: Detection Accuracy versus Classifiers working points, for Random (left) and Optimized (right) surveys deliveries.}\label{fig:TM_pc_others}
\end{figure}

\begin{figure}[t]
\centering
\subfigure[RD strategy, $10k$ users]{\label{fig:10k RD EM}
    \includegraphics[width=0.45\linewidth]{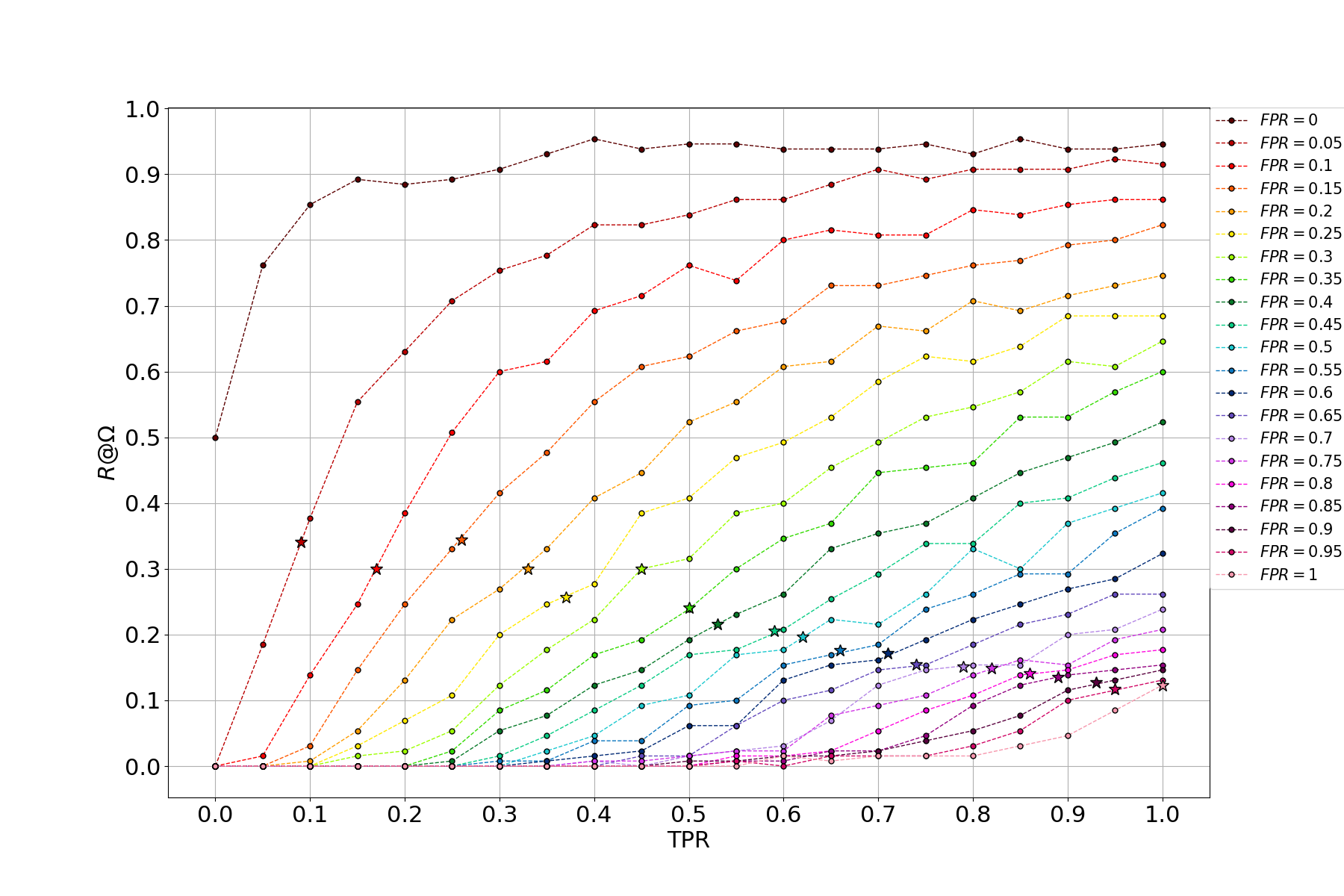}}
\subfigure[OD strategy, $10k$ users]{\label{fig:10k OD EM}
   \includegraphics[width=0.45\linewidth]{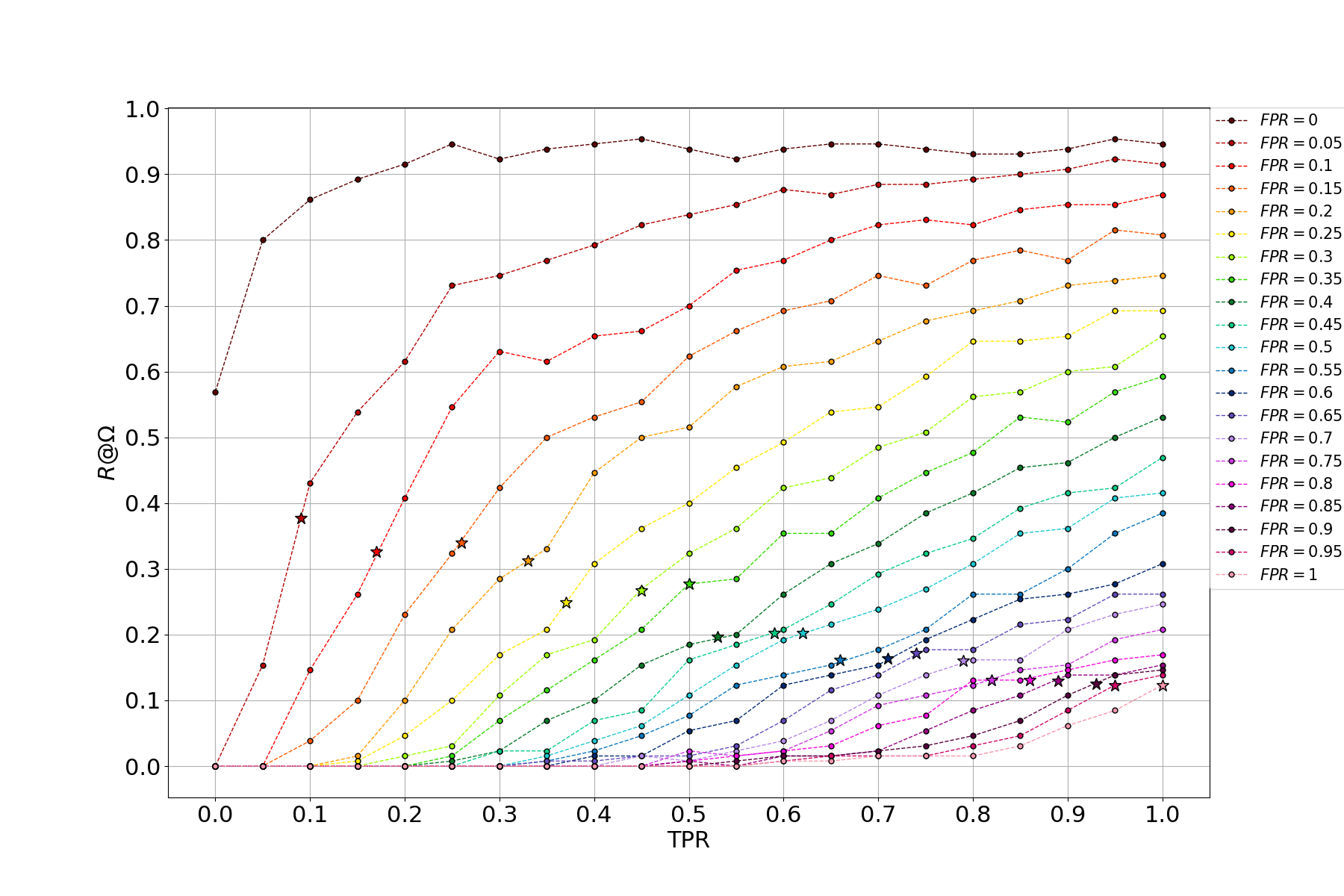}}
\subfigure[RD strategy, $1k$ users]{\label{fig:1k RD EM}
    \includegraphics[width=0.45\linewidth]{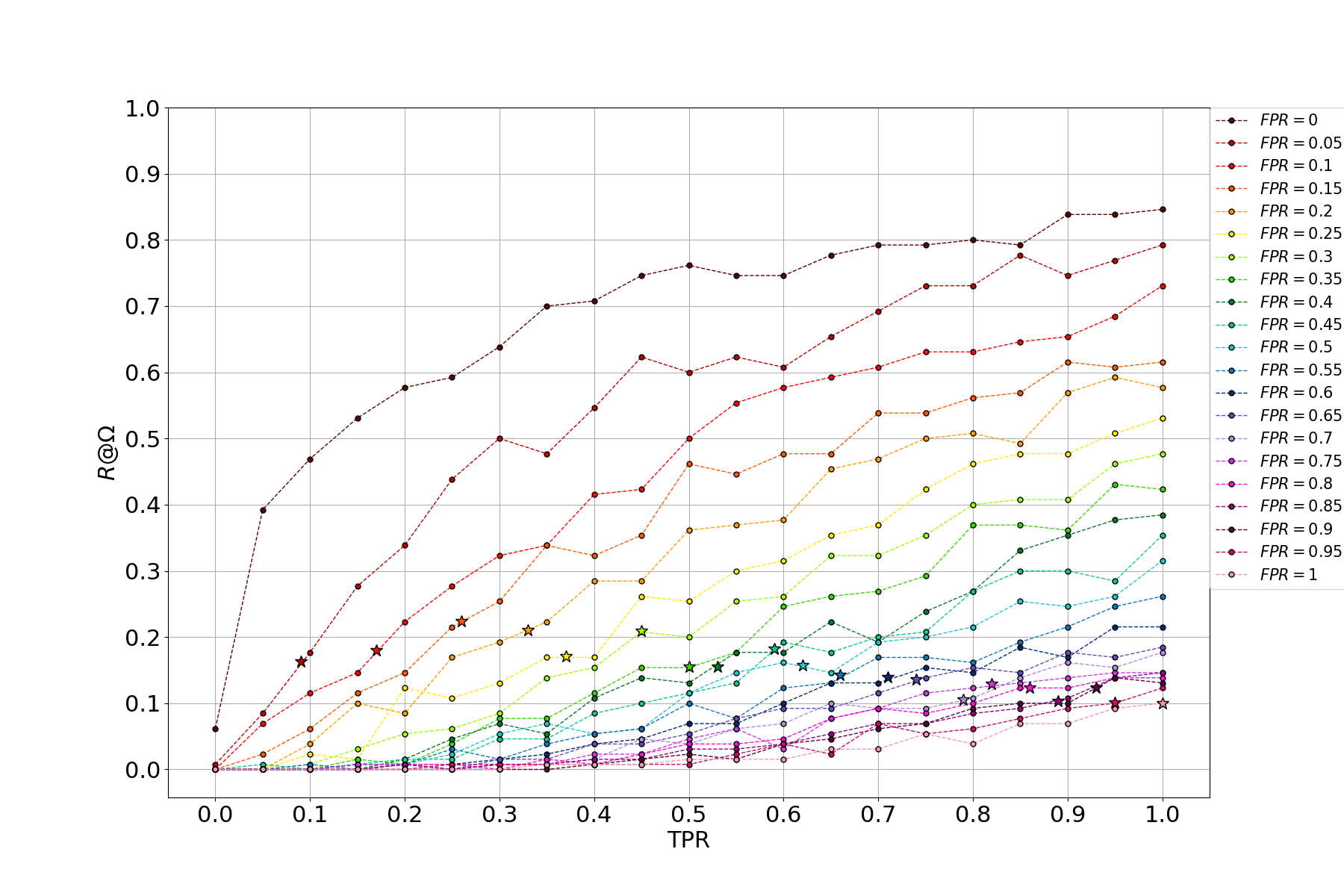}}
\subfigure[OD strategy, $1k$ users]{\label{fig:1k OD EM}
   \includegraphics[width=0.45\linewidth]{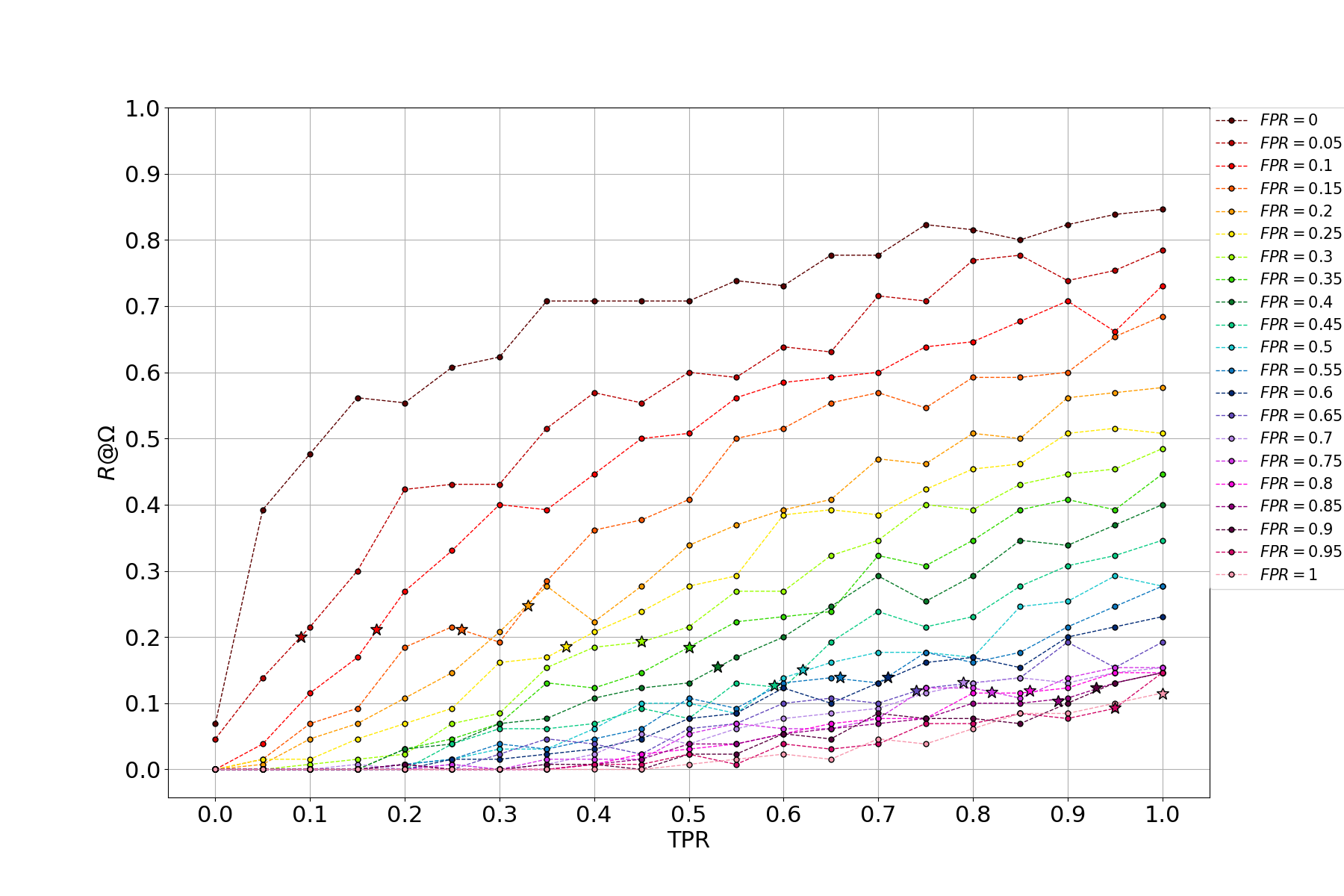}}
\caption{Scenario S2, 10k and 1k Users: Detection Accuracy versus Classifiers working points, for Random (left) and Optimized (right) surveys deliveries.}\label{fig:EM_pc_others}
\end{figure}

It is worth analyzing the best performance $R\textsubscript{C}@\Omega$ achievable by a realistic user satisfaction classifier, such as the one we proposed in \cite{pimpinella2019towards}. We plot its performance points (FPR\textsubscript{C}, TPR\textsubscript{C}) as coloured stars in Figures \ref{fig:TM_pc_100k}, \ref{fig:EM_pc_100k}, \ref{fig:TM_pc_others} and \ref{fig:EM_pc_others}.
For the sake of clarity, we summarise in Tables~\ref{tab:BA_S1} and~\ref{tab:BA_S2} the best values of $R\textsubscript{C}@\Omega$ achievable in all tested scenarios, and we compare it with $R\textsubscript{gt}@\Omega$, the best performance obtained leveraging only the available GT user satisfaction. We observe that $R\textsubscript{gt}@\Omega$ corresponds to the performance of a classifier that predicts each non-GT user as satisfied (i.e., FPR=0 and TPR=0), since satisfied users do not contribute to the ranking score. Therefore, $R\textsubscript{gt}@\Omega$ corresponds to the top-left brown point of a given performance cloud.

\begin{figure}[t]
\centering
\subfigure[$R\textsubscript{gt}@\Omega$ (solid lines) and $R\textsubscript{C}@\Omega$\newline (dashed lines) versus GT users per\newline network site, for RD (red lines) and\newline OD (orange lines) strategies.]{\label{fig:TM_tradeoff} \includegraphics[width=0.45\columnwidth]{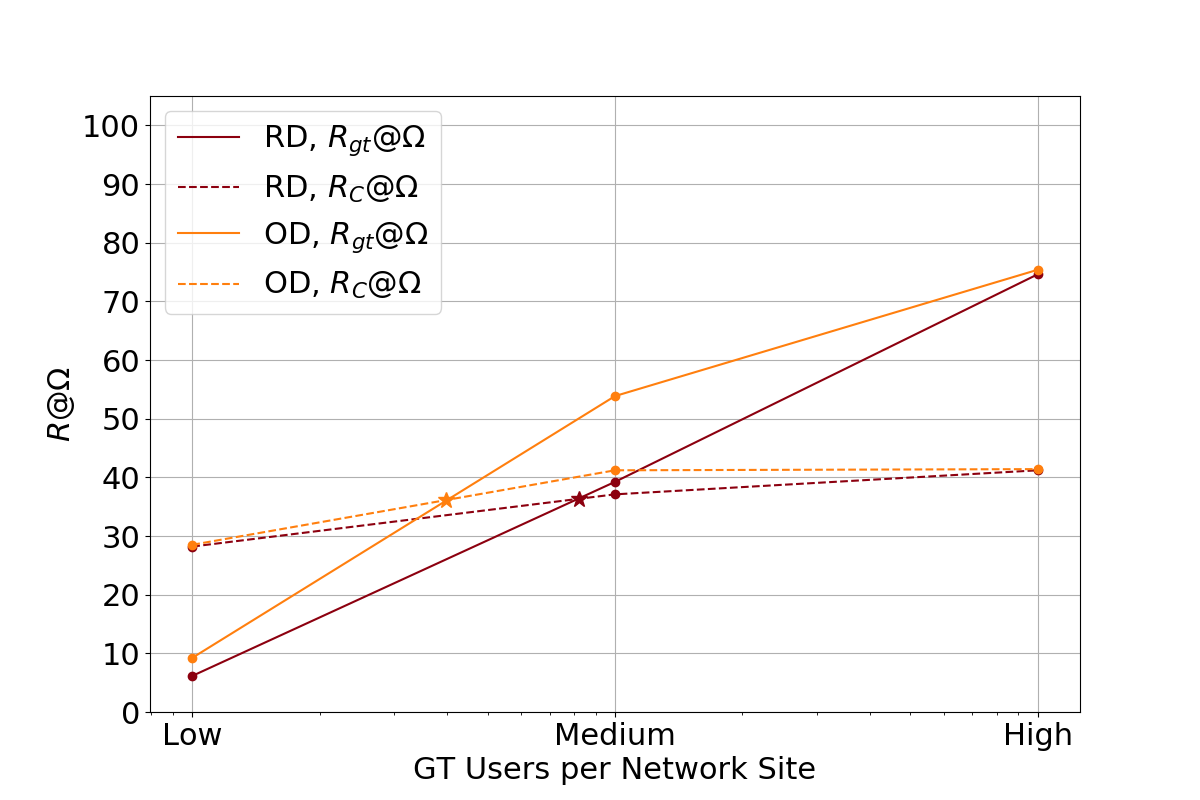}}
 \subfigure[Average network coverage versus GT\newline users per network site, for RD (red lines)\newline and OD (orange lines) strategies.]{\label{fig:TM_coverage} \includegraphics[width=0.45\columnwidth]{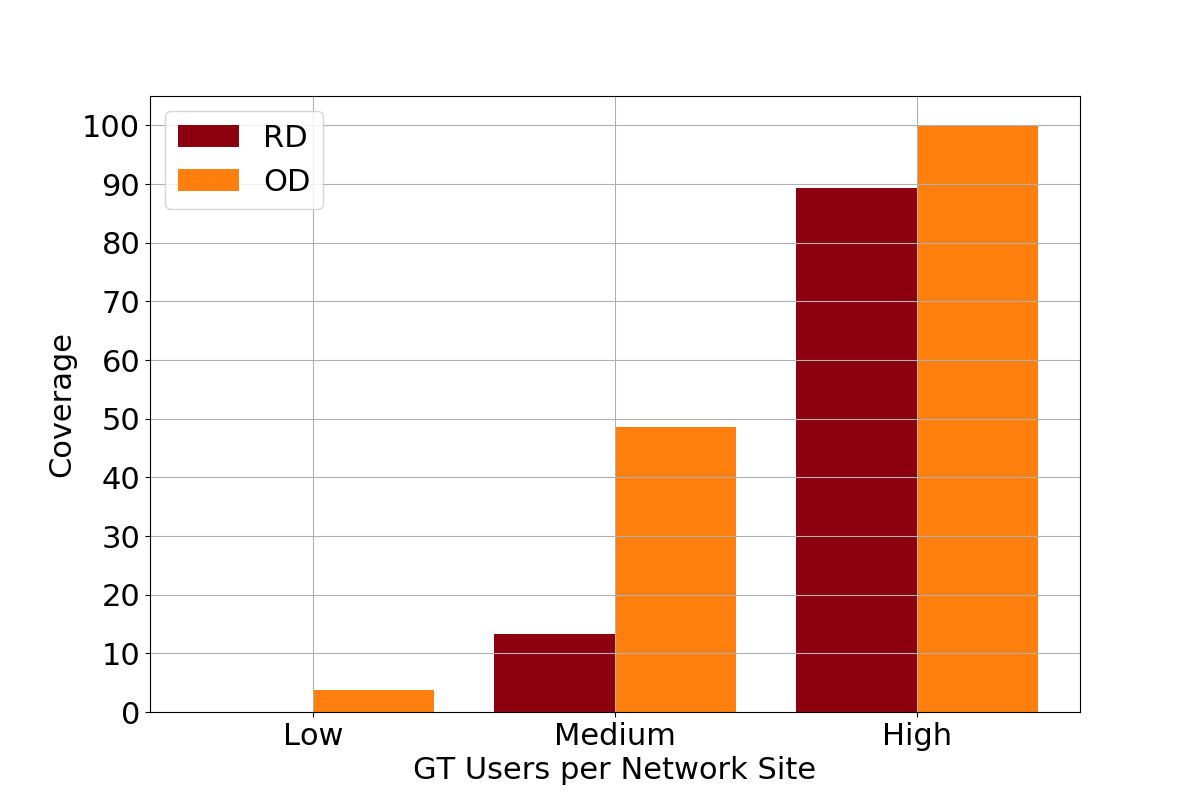}}
\caption{S1 Mobility Scenario.}
\label{fig:TM_results}
\end{figure}

\begin{figure}[t]
\centering
\subfigure[$R\textsubscript{gt}@\Omega$ (solid lines) and $R\textsubscript{C}@\Omega$\newline (dashed lines) versus GT users per\newline network site, for RD (red lines) and\newline OD (orange lines) strategies.]{\label{fig:EM_tradeoff} \includegraphics[width=0.45\columnwidth]{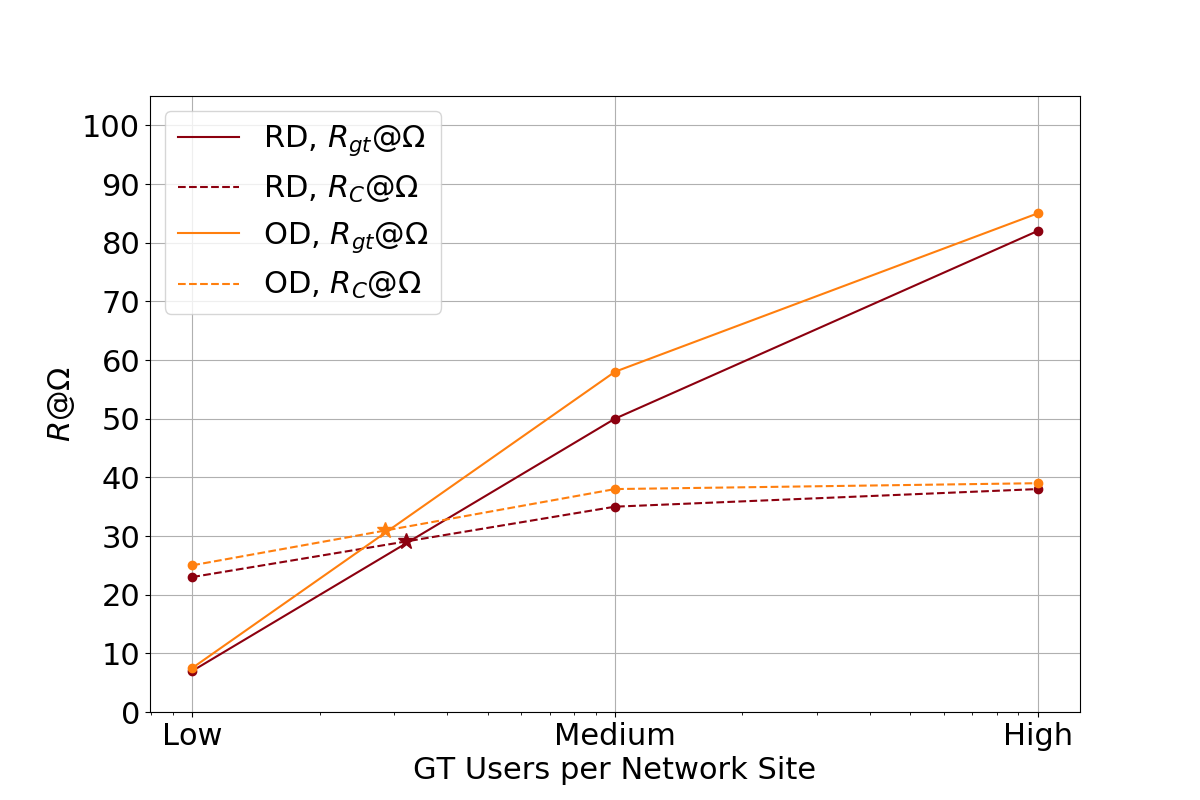}}
\subfigure[Average network coverage versus GT\newline users per network site, for RD (red lines)\newline and OD (orange lines) strategies.]{\label{fig:EM_coverage} \includegraphics[width=0.45\columnwidth]{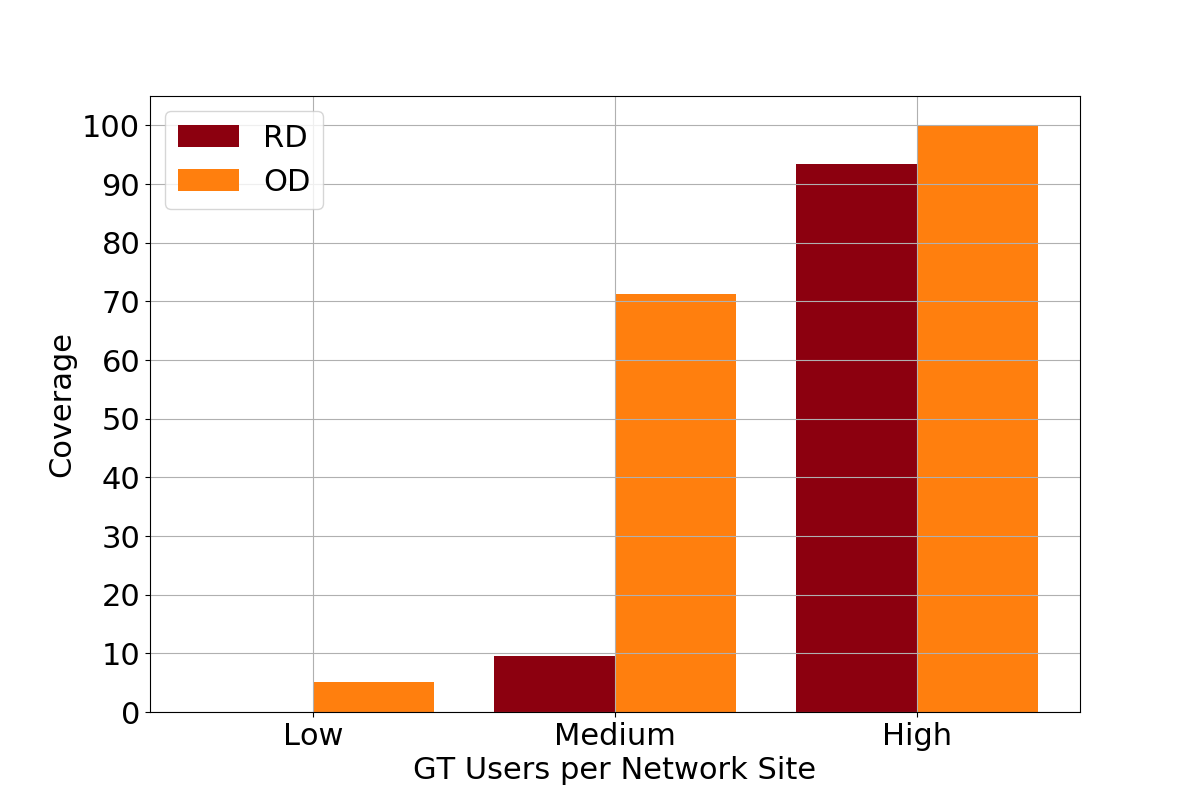}}
\caption{S2 Mobility Scenario.}
\label{fig:EM_results}
\end{figure}

As one can see from Table \ref{tab:BA_S1}, we observe that for high density of GT users $R\textsubscript{gt}@\Omega$ is $33\%$ and $40\%$ higher than $R\textsubscript{C}@\Omega$, for RD and OD strategy respectively. Similarly, in Table \ref{tab:BA_S2}, $R\textsubscript{gt}@\Omega$ is $42\%$(RD) and $45\%$ (OD) higher than $R\textsubscript{C}@\Omega$. In the case of medium GT density, we observe a reduction of the recall gaps for both mobility scenario. For what concerns S1 the recall improves at maximum by $12\%$ (OD), while in the case of S2,$R\textsubscript{gt}@\Omega$ is still $15\%$ (RD) and $19\%$ (OD) better than $R\textsubscript{C}@\Omega$. The situation changes when considering low density of GT users. In this case, we observe that for both mobility scenarios it is better for the operator to leverage the classifier $f_C(\cdot)$ for detecting under-performing sites in the network. In fact, $R\textsubscript{C}@\Omega$ is more than $20\%$ better than $R\textsubscript{gt}@\Omega$ for both delivery strategies in S1 mobility scenario while it is $16\%$ (RD) and $17\%$ (OD) better in scenario S2. 
Such results are also illustrated in Figure \ref{fig:TM_tradeoff} and \ref{fig:EM_tradeoff}. From such figures it is clear that, when using the binary classifier proposed in \cite{pimpinella2019towards}, there exists a critical GT density (represented with a colored star) above which satisfaction prediction becomes detrimental in terms of detection performance. Comparing \ref{fig:TM_tradeoff} with \ref{fig:EM_tradeoff} we also observe that the critical density depends also on the characteristics of user mobility in the network. Since an operator is able to evaluate both the actual GT density available and the users mobility in its own network, it can also take a decision on wheter or not to predict user satisfaction. This answers to \ref{Q4}. 

\subsubsection{Random vs Optimized delivery}
Finally, we discuss the obtained results in order to find an answer for \ref{Q5}. We observe from Figure \ref{fig:TM_tradeoff} and \ref{fig:EM_tradeoff} that the OD strategy always outperform the Random Delivery strategy. Moreover, using the OD strategy has the effect of moving the critical points (yellow star) towards lower densities compared to the RD strategy. The reason of such a better performance is clearly due to the higher coverage that the OD strategy is able to reach. Figures \ref{fig:TM_coverage} and \ref{fig:EM_coverage} show the network coverage for the different scenarios: as one can see, the OD strategy allows to greatly increase the network coverage at different GT users density, which in turns impact on the achievable $R@\Omega$. However, we recall that in case of the OD strategy the operators may need to put in place incentive strategies for receiving the answers from the users selected by the optimization problem.


\section{Related Works}\label{sec:sota}

Many works in literature recognize the importance for cellular operators to monitor service levels at end hosts such to better understand which network events hamper users experience~\cite{faggiani2014smartphone,ren2015exploiting,tong2017research,choffnes2010crowdsourcing,boz2019mobile}. On the one hand, the computational power embedded in today's mobile devices let them be a powerful means for data collection, that can be then processed by the operators for diverse purposes~\cite{faggiani2014smartphone,ren2015exploiting}. On the other hand, the analysis of users experiences in the network and of their corresponding subjective perceptions have become a fundamental benchmark for network operators, which often adopt crowdsourcing strategies to monitor and collect both objective and subjective users side information~\cite{tong2017research,choffnes2010crowdsourcing}. In fact, Quality of Experience (QoE) models can be very helpful to quantify the relationship between users experience and network quality of service~\cite{boz2019mobile}, considering that the more users share the same perception about similar network events the more likely those events share similar QoS characteristics~\cite{choffnes2010crowdsourcing}.

A common way for network operators to collect users QoE evaluations is to issue satisfaction surveys where the customers are asked what is their likelihood regarding the experienced mobile services. Then, operators can for example leverage the collected QoE feedbacks to plan actions to minimize the churn-rate of their customers,i.e., the percentage of customers who stop their contributions and move to a different operator due to unsatisfactory service~\cite{huang2015telco,swetha2018evaluation,pimpinella2019towards}. In ~\cite{hossfeld2013best} the authors identify four main categories that influence the satisfaction of cellular users, namely \textit{context}, user \textit{profile}, \textit{system} and \textit{content}. The \textit{context} considers factors like the purpose of using the service, the user's cultural background and the environment in which the user uses the service while the user \textit{profile} considers individual psychological factors and memory. Finally, \textit{system} and \textit{content} address respectively technical influence factors (such as device-related problems) and resolution/format related issues. 

However, a common problem found by cellular operators to assess the QoE of their customers through crowdsourced surveying campaigns is that few users usually respond to satisfaction surveys. To counteract this problem, usually operators implement techniques to estimate or predict users QoE feedbacks from objective network mesurements. Many works in literature tackle the (complex) issue of predicting users QoE in mobile networks, differentiating between short-term (~\cite{casas2017predicting,casas2016next,nam2016qoe,de2012quantifying,hossfeld2011quantification}) and long-term (~\cite{huang2015telco,swetha2018evaluation,pimpinella2019towards,hossfeld2011memory} users experiences. 
On the one hand, a short-term network experience refers to the case in which a user is first requested to interact with a mobile application under variable (and manually controlled) QoS network levels and secondly asked to provide a QoE evaluation of the experience. In~\cite{casas2017predicting}, authors use in-smartphone measurements to feed several ML algorithms and predict users cellular users QoE with respect to several mobile applications. Leveraging a dataset comprised of 30 users, which were requested to watch short videos and give QoE feedbacks for each session, they obtain 91\% and 98\% accuracy on users feedbacks and service acceptability level respectively. A similar work is described in~\cite{casas2016next}, where the authors conduct both lab tests and on field trials to analyse the impact of many network related features (e.g., bandwidth, latency, etc.) on users QoE of common mobile applications. Interestingly, in both~\cite{nam2016qoe,hossfeld2011quantification} the authors show that users QoE of video streaming applications is primarily influenced by the frequency and duration of stalling events, i.e., the longer the video playback re-buffering time the more likely the user will stop watching it. Similarly, authors in~\cite{de2012quantifying} recognize from subjective users QoE assessments that i) long video re-buffering and loading time are perceived as highly disturbing by the users and ii) fluent playbacks are preferred with respect to other video-related service indicators (such as resolution, frame rate or bit-rate). In other words, the longer the users experience disturbing network events the more likely their QoE will decrease.

On the other hand, the prediction of long-term users satisfaction is a much more challenging task to address. This is because a long-term user experience in a mobile network composes of many and different network events which together influence his/her QoE of the received cellular service. This means that users memory plays an important role in long-term QoE assessment processes, as discussed in~\cite{hossfeld2011memory}. Memory effects are also investigated in~\cite{huang2015telco}, where the authors leverage a large volume of network data regarding the experience of users in the network of one of the biggest mobile operator in China over several months, with the final aim of implementing a churn prediction system. They also integrate the prediction system in a closed loop automatic retention mechanism, with the aim of both acquiring new customers while retaining potential churners. Their results show that such a system improved the recharge rate of potential churners of more than 50\%. With the same aim, in~\cite{swetha2018evaluation} authors introduce a modified random forest algorithm able to estimate a cellular customer's churn rate yielding an AUC value of $91.5\%$. Similarly, in~\cite{pimpinella2019towards} the authors correlate user-side network measurements with corresponding QoE feedbacks to train several ML algorithms and predict users satisfaction about network coverage and video streaming services. Moreover, considering that different users visit several network areas/elements and that the same area/element is usually visited by many different users, they point out that i) the information about ground truth and predicted users QoE feedbacks together with network measurements data can be used to recognize what in the network causes users dissatisfactions and ii) the impact of misclassification errors on such process could be somehow reduced when users QoE information are grouped on a single network area/element. 

To conclude, considering that users QoE feedbacks are by definition subjective, an important issue regards the \textit{reliability} of users answers to satisfaction surveys. Many works~\cite{raykar2010learning, karger2011iterative, shah2015approval, hossfeld2013best} show that gathering reliable information from a crowd is a very challenging task. In~\cite{raykar2010learning} the authors give a probabilistic approach for supervised learning in a situation when there are possibly noisy replies collected from multiple users and there are no absolute gold standards (i.e. standard questions used to evaluate the level of reliability of experts). Similarly, authors in~\cite{karger2011iterative} propose an iterative algorithm for deciding best survey allocation and calculating a weighted estimate of the correct survey answer. Interestingly, in~\cite{shah2015approval} it is shown how an incentive-compatible compensation algorithm together with approval-voting mechanisms successfully convert a significant fraction of incorrect answers to correct replies at the price of little increase in net expenditures.

\section{Concluding Remarks}\label{sec:conclusion}
In this work we considered the process of crowdsourcing-based network monitoring, which may be used by cellular operators to detect problems in their network on the basis of users satisfaction feedbacks. We observe that several aspects need to be considered by an operator that decides to leverage such an approach. On the one hand, the heterogeneous reactions of users to service issues can hamper the detection of malfunctions in the network. On the other hand it is not trivial to understand which network site is the main responsible of a user feedback, considering that each user visits many network sites for different amounts of time. Moreover, often very few users participate in the crowdsourcing process, thus forcing the operator to implement ML algorithms able to predict users satisfaction on the basis of objective measurements, in order to enlarge the knowledge base usable for monitoring purposes. This introduces a further aspect, which regards the impact of prediction errors on the detection of issues in the network. For all these reasons, we implemented a simulation framework that can be used by a cellular operator to analyse the application of a crowdsourcing-based network monitoring process in different realistic scenarios and investigate the related aspects. From the results we obtained, the following conclusions can be drawn:
\begin{itemize}

	\item Under the reasonable assumption that users satisfaction depends on the performance of the visited network sites, it is possible for a network operator to rank/detect malfunctioning sites leveraging users satisfaction feedbacks with good detection performance (as shown in Figure \ref{fig:RvsTTAvsUT});
	
	\item The detection process works regardless of the satisfaction profile of the visiting users, which in this work is represented by a random variable that controls users tolerance to bad network events. In particular, Figure \ref{fig:RvsTTAvsUT} shows the robustness of the process with respect to the average users tolerance $\mu$ and to the threshold $\xi$;
	
	\item If a binary classifier $f(\cdot)$ is included in the detection process, working at low FPR rather than high TPR is more rewarding in terms of detection performance (as observed in Figures \ref{fig:TM_pc_100k}, \ref{fig:EM_pc_100k}, \ref{fig:TM_pc_others} and \ref{fig:EM_pc_others});
	
	\item When the coverage of the network ensured by GT users is low, it is convenient for an operator to leverage a ML classifier to predict the satisfaction of non-GT users such to augment the knowledge base usable for detection purposes. Moreover, this is true even when the classification performance of the ML classifier are modest, as shown in Figures \ref{fig:TM_pc_100k}, \ref{fig:EM_pc_100k}, \ref{fig:TM_pc_others} and \ref{fig:EM_pc_others}. Conversely, for high coverage values, it is better for the operator to rely only on GT users for detecting under-performing sites in the network. These results are summarised in \ref{fig:TM_results} and \ref{fig:EM_results}; 

	\item The implementation of delivery strategies that optimally allocate satisfaction surveys to users such as to maximize the network coverage increases the detection performance, as observed in Figures \ref{fig:TM_results} and \ref{fig:EM_results};
\end{itemize}

We believe these observations can be useful for a network operator willing to adopt crowdsourcing-based network monitoring.

\balance

\appendix
\section{Maximum Coverage Problem Formulation} \label{sec:ILP}
In this Section, we describe the optimization problem that can be run by the Survey Delivery block to optimize the delivery of the surveys in order to maximize the network coverage. The optimization problem is the budgeted version \cite{khuller1999budgeted} of a family of well-known problems known as \textit{Maximum Coverage} (MC) problems. Given a collection $\mathcal{S}$ of items with associated costs defined over a domain of weighted elements and a budget $B$, the (budgeted) MC problem aims to find a subset $\mathcal{S}' \cup \mathcal{S}$ such that the total cost of items in $\mathcal{S}'$ does not exceed B and the total weight of the \textit{covered} elements is maximized. In our case, we want to deliver satisfaction surveys to users such that the number of \textit{covered} network sites is maximised, where a network site is covered if (i) it is visited by at least $n$ users and (ii) each user spends more than $\xi$ percentage of its own time in the site.
Table \ref{tab:param_mcp} summarizes the parameters that are leveraged by the optimization program. Let:

\begin{itemize}
	\item $x_i$ be a binary variable equals to 1 if a satisfaction survey is delivered to user $i$ and zero otherwise;
	\item $c_j$ be a binary variable equals to 1 if network site $j$ is \textit{covered} and zero otherwise;
	\item $h_{i,j}$ be a binary association variable which equals 0 if the time that user $i$ has spent in site $j$ is lower than $\xi T$ (i.e., if it is not sufficient for coverage), while it can be both 0 or 1 otherwise. 
\end{itemize} 

\begin{table}[t]
\centering
\caption{Parameters considered in the MC problem.}
\label{tab:param_mcp}
\begin{tabular}{|c|c|c|}
\hline
\multicolumn{1}{|l|}{\textbf{Parameter}} & \textbf{Definition} \\ \hline
$M$ & Number of Network Sites 
\\ \hline
$\mathcal{J}$  & Set of Network Sites
\\ \hline
$\mathcal{I}=\{1, .., N\}$ &  Set of Users 
  \\ \hline
$B$  & GT Users Budget 
\\ \hline
$T$  & Time Horizon 
\\ \hline
$t_{i,j}$ & User-site visit time 
\\ \hline
$\xi$ & \begin{tabular}{c} Percentage of time a user\\ needs to spend in a site for covering it\end{tabular}  
\\ \hline
$n$ & \begin{tabular}{c} Minimum number of visitors to consider a site covered\end{tabular} 
\\ \hline
\end{tabular}
\end{table}

Under these definitions, we propose an Integer Linear Programming (ILP) formulation for our version of the budgeted MC problem as it follows:
\begin{align}
\max_{x_i} &\quad& &\sum_{j \in \mathcal{J}}c_j \label{eq:obj_func} & \\
\text{subject to: } \nonumber & \\
&\quad&  &\sum_{i \in \mathcal{I}} x_{i} \leq B \label{eq:budget} & \\
\vspace{5mm}
&\quad&  &t_{i,j} \geq \xi \cdot T \cdot h_{i,j} &\quad&\forall{(i,j)} \in \mathcal{I}\times\mathcal{J} \label{eq:mintime} & \\
\vspace{5mm}
&\quad&  &\sum_{j \in \mathcal{J}} h_{i,j} \geq 1-M \cdot (1-x_{i})&\quad&\forall{i} \in \mathcal{I} \label{eq:user_act1}  & \\
\vspace{5mm}
&\quad& &\sum_{j \in \mathcal{J}} h_{i,j}< 1+M \cdot x_{i}&\quad&\forall{i} \in \mathcal{I} \label{eq:user_act2} & \\
\vspace{5mm}
&\quad& &\sum_{i \in \mathcal{I}} h_{i,j}\geq n-M \cdot(1-c_{j})&\quad&\forall{j} \in \mathcal{J} \label{eq:site_cov1} &\\
\vspace{5mm}
&\quad& &\sum_{i \in \mathcal{I}} h_{i,j}<n+M \cdot c_{j}&\quad&\forall{j} \in \mathcal{J} \label{eq:site_cov2}
\end{align}
Equation \ref{eq:obj_func} represents the objective function, which aims at maximizing the number of distinct covered sites. Constraint \ref{eq:budget} limits the number of distinct users answering a survey to be lower or equal then the GT users budget $B$. Note that for large population sizes $N$ the number of users which ensures full coverage could be smaller than the budget. Constraint \ref{eq:mintime} controls the minimum time needed for a user $i$ to contribute to the coverage of site $j$. Note that while the variable $h_{i,j}$ is forced to be 0 when the time spent by user $i$ is not sufficient to be a covering visitor of site $j$, it is not constrained to be 1 if the coverage condition is met, so that the solver can decide which user is more convenient to activate to maximize the objective function. Constraints \ref{eq:user_act1} and \ref{eq:user_act2} control the selection of a generic user $i$ for the delivery of the survey (i.e. the activation of user $i$), which arises from the activation of the corresponding time variable $h_{i,j}$ for at least one of the visited network sites. In particular, \ref{eq:user_act1} forces the variable $x_{i}$ to be 0 in the case in which the summation on the left is 0, whereas \ref{eq:user_act2} forces the same variable to be 1 in the case in which the corresponding summation is strictly greater than 0. Note that when \ref{eq:user_act1} forces $x_{i}$ to be 0, then \ref{eq:user_act2} deactivates, while the opposite happens when \ref{eq:user_act2} forces $x_{i}$ to equal 1. Finally, constraints \ref{eq:site_cov1} and \ref{eq:site_cov2} set the requirements for considering a site as covered and work similarly to constraints \ref{eq:user_act1} and \ref{eq:user_act2}. 


\bibliographystyle{elsarticle-num-names}
\bibliography{sample.bib}







\end{document}